\documentclass[11pt]{article}
\usepackage{amsmath,amssymb,graphicx}
\usepackage{hyperref}
\usepackage{cite}

\usepackage{multicol,color,longtable}

\definecolor{darkred}{rgb}{0.5,0,0.5}
\hypersetup{pdfborder={0 0 0},colorlinks=true,urlcolor=darkred,citecolor=blue,linkcolor=darkred}

\makeatletter

\@addtoreset{equation}{section}
\makeatother

\newcommand{\nn}{\nonumber}

\newcommand{\ints}{\mathbb{Z}}
\newcommand{\reals}{\mathbb{R}}

\newcommand{\lb}{\left[}
\newcommand{\rb}{\right]}

\newcommand{\ld}{\left\{}
\newcommand{\rd}{\right\}}

\newcommand{\cS}{\mathcal{S}}
\newcommand{\cH}{\mathcal{H}}
\newcommand{\cC}{\mathcal{C}}

\newcommand{\ta}{{\tt{a}}}
\newcommand{\tb}{{\tt{b}}}

\newcommand{\Ga}{\Gamma}
\newcommand{\veps}{\varepsilon}
\newcommand{\Gt}{\tilde\Gamma}
\newcommand{\vepst}{\tilde\varepsilon}
\newcommand{\si}{\sigma}

\newcommand{\E}{{\rm{E}}_{10}}

\newcommand{\tone}{{\tilde{1}}}
\newcommand{\ttwo}{{\tilde{2}}}
\newcommand{\tthree}{{\tilde{3}}}
\newcommand{\Tr}{\textrm{Tr}\,}

\newcommand{\Aof}[1]{A_{#1}}
\newcommand{\Eof}[1]{E_{#1}}
\newcommand{\Jof}[1]{J_{#1}}

\newcommand{\Pof}[1]{P_{#1}}
\newcommand{\Phof}[1]{\hat{P}_{#1}}
\newcommand{\Qof}[1]{Q_{#1}}
\newcommand{\Piof}[1]{\Pi_{#1}}
\newcommand{\Pihof}[1]{\hat{\Pi}_{#1}}
\newcommand{\kof}[1]{{\rm k}_{#1}}
\newcommand{\jof}[1]{{\rm j}_{#1}}
\newcommand{\pihat}{\hat{\pi}}

\newcommand{\xnull}{\mathbf{x}_0}

\begin{document}

\thispagestyle{empty}

{\flushright {\tt AEI-2014-057}\\[16mm]}

\begin{center}
{\LARGE \bf \sc Canonical Structure of the $\E$ Model\\[2mm] and Supersymmetry}\\[8mm]

\vspace{8mm}
\normalsize
{\large  Axel Kleinschmidt${}^{1,2}$, Hermann Nicolai${}^1$ and Nitin K. Chidambaram${}^3$}

\vspace{10mm}
${}^1${\it Max-Planck-Institut f\"{u}r Gravitationsphysik (Albert-Einstein-Institut)\\
Am M\"{u}hlenberg 1, DE-14476 Potsdam, Germany}
\vskip 1 em
${}^2${\it International Solvay Institutes\\
ULB-Campus Plaine CP231, BE-1050 Brussels, Belgium}
\vskip 1 em
${}^3${\it Indian Institute of Technology Madras, Department of Physics\\
Chennai -  600036, India}

\vspace{20mm}

\hrule
\vspace{5mm}
\begin{tabular}{p{12cm}}
{\footnotesize 
A coset model based on the hyperbolic Kac--Moody algebra $\E$ has been conjectured 
to underly eleven-dimensional supergravity and M theory.
In this note we study the canonical structure of the bosonic model for finite- and infinite-dimensional groups. In the case of finite-dimensional groups like $GL(n)$ we exhibit a convenient set of variables with Borel-type canonical brackets.  The generalisation to the Kac--Moody case requires a proper treatment of the imaginary roots that remains elusive. As a second result, we show that the supersymmetry constraint of $D=11$ supergravity can be rewritten in a suggestive way using $\E$ algebra data. Combined with the canonical structure, this rewriting explains the previously observed association of the canonical constraints with null roots of $\E$.  We also exhibit a basic incompatibility between local supersymmetry and the $K(\E)$ `R symmetry',  that can be traced 
back to the presence of imaginary roots and to the unfaithfulness of the spinor 
representations occurring in the present formulation of the $\E$ worldline model,
and that may require a novel type of bosonisation/fermionisation for its resolution.
This appears to be a key challenge for future progress with $\E$.
}
\end{tabular}
\vspace{5mm}
\hrule

\end{center}

\setcounter{page}{0}
\newpage

\section{Introduction}

The conjectured $\E$ symmetry of the M Theory completion of
$D=11$ supergravity~\cite{Damour:2002cu} applies to both the bosonic and the fermionic sector. The one-dimensional spinning $\E$ model constructed and analysed in~\cite{Damour:2005zs,deBuyl:2005mt,Damour:2006xu,deBuyl:2005zy} has manifest symmetry under the hyperbolic Kac--Moody group $\E$ and its dynamics have been shown to match exactly the $D=11$ dynamics at the non-linear level, when both are suitably truncated. However, it has so far proved impossible to remove the truncation of this correspondence, one central obstacle being a dichotomy between the bosonic and fermionic variables on the $\E$ side. Whereas the bosonic variables are described in terms of {\em infinitely} many coordinates of the infinite-dimensional coset space 
$\E/K(\E)$, the fermionic variables are described by {\em finitely} many components of a 
finite-dimensional (unfaithful) spinor representation of $K(\E)$~\cite{Damour:2006xu}. 
This dichotomy is also reflected in the fact that the one-dimensional $\E$ model cannot be fully supersymmetric on its worldline, since in its presently known form it pairs an infinite number of bosonic with a finite number of fermionic degrees of freedom.

In view of the fact that a detailed understanding of supersymmetry has 
often been central in advances regarding the structure of hidden symmetries, 
%see for instance~\cite{deWit:1986mz,Godazgar:2014eza}, 
we initiate in this note a 
more detailed study of the worldline supersymmetry in the $\E$ context. Though we will not be able to present a new supersymmetric $\E$ model, our results bring the obstacles in the current formulation to the front and we hope they can serve as a first step to resolving the issues 
both in the physics and the mathematics associated with constructing a model that
fully accommodates both supersymmetry and $K(\E)$ symmetry. In fact,  progress towards 
solving the outstanding problems may well require some novel kind of 
bosonization/fermionization, and thus also involve quantisation in a crucial way. This is 
not only because the distinction between bosons and fermions becomes fluid  in low 
dimensions and thus also in the (one-dimensional) worldline model, but also because the
very meaning of what is a space-time boson and what is a space-time fermion, and hence 
also the ultimate relevance of {\em space-time} supersymmetry (as opposed to worldline
supersymmetry), must be questioned 
in the context of emergent space-time scenarios. The present results can be viewed as
a first step in this direction; in particular we identify the proper canonical variables 
on the bosonic side that couple naturally to the fermions, and hence will be an
essential ingredient in approaching quantisation of the worldline model. We note 
that in the context of string theory the emergence of space-time fermions from 
bosonic fields was already suggested long ago in~\cite{Casher:1985ra},
and the relation of this construction to Kac--Moody algebras was 
discussed more recently in~\cite{Englert:2008ft}. In the context of maximal 
supergravity in two dimensions (where $K(\E)$ is replaced by $K({\rm E}_9)$),
it was already pointed out in \cite{Nicolai} that the associated linear system effectively
constitutes a bosonisation of the supergravity fermions, especially in view of previous work
in \cite{Witten,GNO}.

Our main tool is the detailed analysis of the canonical structure of one-dimensional coset models, starting with purely bosonic systems based on a coset $G/K$. We will exhibit explicitly a set of  variables that makes the algebraic structure completely manifest and we propose that these variables are therefore also an appropriate starting point for quantum considerations extending the reduced quantum cosmological billiards of~\cite{Kleinschmidt:2009cv} that should eventually
lead to an implementation of the Wheeler-DeWitt equation for the full theory.
For the case $\E/K(\E)$ our arguments remain somewhat formal since an explicit parametrisation of the group $\E$ similar to the one used in the proof for finite-dimensional $G$ is not available. 
Denoting the velocity type variables as $\Pof{\alpha}^r$, where $\alpha >0$ is a positive
root of the $\E$ Borel algebra and $r$ labels an orthonormal basis of elements in the 
root space associated to the root $\alpha$ (this extra label is only required for imaginary 
roots), we will in particular argue, and {\em prove} for finite-dimensional $G$,
that the canonical commutation relations of 
the $\Pof{\alpha}^r$ are exactly those of the $\E$ algebra itself.

The bosonic expressions have to be completed by fermionic ones and in section~\ref{sec:fermions} we then look at $D=11$ supergravity \cite{Cremmer:1978km}.
A rewriting of the supersymmetry constraint, inspired by recent studies in quantum supersymmetric cosmology in relation to Kac--Moody symmetries~\cite{Damour:2013eua,Damour:2014cba}, suggests a very simple underlying algebraic formulation. 
We will here restrict attention to terms {\em linear in the fermions}, as the consideration
of higher order fermionic terms does not affect our main conclusions.\footnote{By contrast, 
the supersymmetric Bianchi-type 
model recently (and impressively) analysed in~\cite{Damour:2013eua,Damour:2014cba} 
does retain all higher order fermionic terms, and is thus fully supersymmetric also 
at the quantum level, but with only partial manifestations of $K({\rm AE}_3)$ symmetry.}
With every root $\alpha$ of $\E$ one can associate an element $\Gt(\alpha)$ of the $SO(10)$ Clifford algebra and a polarisation of the fermionic field $\phi(\alpha)$.  In \cite{Damour:2005zs} the 
supersymmetry constraint was analysed {\em to linear order} in fermions and shown 
to take the schematic form
\begin{align}
\label{PodotPsi}
\mathcal{S} = P \odot \Psi 
\end{align}
where $P$ stands for the infinite component coset velocity of the $\E$ coset
space model, and $\Psi$ for the {\em finite}-dimensional unfaithful spinor representation.
The symbol $\odot$ is shorthand for the particular combination of the fermions
and the bosonic coset velocities identified from the canonical supersymmetry
constraint in \cite{Damour:2005zs}. In this paper, we will show how the above 
expression can (again schematically) be transformed into a sum
\begin{align}
\label{Sintro}
\mathcal{S} = \ldots \, + \, \sum_\alpha \Pof{\alpha} \, \Gt(\alpha) \phi(\alpha) +\ldots .
\end{align}
One main goal of this paper will be to explore the validity, and more specifically
the limit of validity, of this expression, and thereby attach
%as will be the derivation of and the dots in the formula, 
a more concrete representation theoretic meaning to the symbol $\odot$. Indeed,
already from the form of (\ref{Sintro}) one may anticipate problems 
when trying to combine supersymmetry with the `R symmetry' $K(\E)$: supersymmetry
requires an {\em  equal} number of bosons and fermions, whereas in (\ref{Sintro}) 
an infinite number of bosonic degrees of freedom is to be paired with a finite number 
of fermionic degrees of  freedom. To be sure, in the actual expression obtained from 
supergravity the above sum contains only finitely many bosonic contributions, as a 
result of  `cutting off' the sum over roots $\alpha$ at level $\ell =3$. Therefore 
the supersymmetry constraint $\cS$ cannot, in its presently known form, be assigned 
to any known representation of $K(\E)$, even though separately, both $P$ and $\phi$ 
do transform properly (although it is not known whether $P$ transforms in an 
{\em irreducible} representation of $K(\E)$).
The novel techniques introduced in this paper will allow us 
to analyse in considerable detail the terms by which the supersymmetry constraint 
fails to transform properly, and to highlight the differences between the finite-dimensional 
and infinite-dimensional cases. Our analysis thus identifies the terms that have to be 
dealt with differently in the construction of a supersymmetric $\E$ model, and we 
offer more comments in the concluding section. There we also explain that the failure to transform covariantly under $K(\E)$ cannot be cured by higher order fermionic terms.

While the exact $D=11$ supersymmetry constraint can be transformed into a {\em truncated} expression of the type above, we thus encounter obstacles when trying to remove
the truncation and to explore what the dots in the above formula could stand for.
The expression above does provide a sensible object for $GL(n,\reals)$ and other 
finite-dimensional groups in the sense that it transforms covariantly as spinor as the 
supersymmetry should, but a similar result is no longer true for $\E$. 
From a more physical perspective,  the mismatch between bosons and 
fermions in the latter case is also reflected in the fact that no 
fermionic analog of the gradient representations has been found so far, thus (so far,
at least) precluding an expansion for the fermions \`a la Belinksi--Khalatnikov--Lifshitz (BKL).

\section{Canonical structure of bosonic worldline coset models}
\label{sec:Can}

In this section, we study the canonical structure of a coset model describing the motion 
of a point particle on a symmetric space $G/K$, with $G$ a split real simple Lie group 
and $K\equiv K(G)$ its maximal compact subgroup. To set the basic notations and conventions, 
we first discuss the case of finite dimensional $G$ where everything is well defined,
and subsequently write down the corresponding expressions for Kac--Moody algebras
and groups. In the latter case, of course, many expressions will remain formal. For previous work on the canonical structure of non-linear $\sigma$-models, see for example~\cite{Matschull:1994vi}.

\subsection{Set-up in the finite-dimensional case}

To begin with, we restrict attention to finite-dimensional and simply-laced Lie group $G$. Then the Lie algebra $\mathfrak{g}=\mathrm{Lie}(G)$ is finite-dimensional and has a system of roots $\alpha\in\Delta=\Delta_-\cup\Delta_+$. The positive roots will also be written as $\alpha>0$, and we designate the Cartan subalgebra by $\mathfrak{h}$. We assume a Cartan--Weyl basis with basis vectors $H_\ta$ and $\Eof{\alpha}$, where $\ta=1,\ldots,\dim \mathfrak{h}$. The commutation relations are~\cite{GoddardOlive}
\begin{subequations}
\label{CWBasisFD}
\begin{align}
\label{CWBasis1}
\lb \Eof{\alpha}, \Eof{\beta} \rb &= \left\{ \begin{array}{cl} c_{\alpha,\beta} \Eof{\alpha+\beta} &\textrm{if $\alpha+\beta\in \Delta$,}\\[2mm]
\alpha^\ta H_\ta  & \textrm{if $\alpha=-\beta$,}\\[2mm]
0 & \textrm{otherwise,}
\end{array}\right.\\[2mm]
\lb H_\ta , \Eof{\alpha} \rb &= \alpha_\ta \Eof{\alpha}.
\end{align}
\end{subequations}
with $c_{\alpha,\beta} = \pm 1$  or $= 0$ for simply laced finite-dimensional Lie algebras.
There is a non-degenerate invariant bilinear form on $\mathfrak{g}$ that
satisfies\footnote{For finite-dimensional simple $\mathfrak{g}$ this is just the appropriately 
normalised matrix trace.} 
\begin{subequations}
\begin{align}\label{Norm1}
\langle \Eof{\alpha} | \Eof{\beta} \rangle&=\left\{\begin{array}{cl} 1&\textrm{if $\alpha=-\beta$},\\[1mm]
0 &\textrm{otherwise},
\end{array}\right.\\[2mm]
\langle H_\ta | H_\tb \rangle &= G_{\ta\tb}.
\end{align}
\end{subequations}
The metric $G_{\ta\tb}$ is positive definite for any simple finite dimensional Lie algebra
$\mathfrak{g}$ but need not be positive definite for {\em non-simple} $\mathfrak{g}$.
The inverse $G^{\ta\tb}$ of $G_{\ta\tb}$ has been used to raise the index in~\eqref{CWBasis1} according to $\alpha^\ta=G^{\ta\tb} \alpha_\tb$ and $\alpha_\ta=\alpha(H_\ta)$. The compact subalgebra $K(\mathfrak{g})\equiv\mathfrak{k}\subset \mathfrak{g}$ is generated by $\kof{\alpha}\equiv\Eof{\alpha}-\Eof{-\alpha}$ with $\alpha>0$ and will be discussed in more detail in section~\ref{sec:KE10}. The structure constants $c_{\alpha,\beta}$ are 
antisymmetric and satisfy standard identities~\cite{GoddardOlive}, in particular
\begin{equation}\label{cab}
c_{\alpha,\beta} =  - c_{\beta,\alpha} = - c_{-\alpha, -\beta} \; \;, \quad
c_{\alpha + \beta, -\beta} = c_{\alpha,\beta}.
\end{equation}

The coset $G/K(G)\equiv G/K$ can be parametrised in a Borel gauge fixed form according to the Iwasawa decomposition $G=KAN$. For finite-dimensional $G$ any element of the coset $G/K$
can thus be written in the form\footnote{Here, we deviate from the standard notation 
  in the cosmological billiards literature (see e.g. \cite{Damour:2002et}) where the diagonal     degrees of freedom are denoted by $-\beta^\ta \equiv + q^\ta$. This is done in order to avoid confusion with the labeling   of the components of the root $\beta$ below.} 
%  $A=\mathrm{diag}(e^{-\beta^1},\ldots,e^{-\beta^n})$. 
\begin{align}
\label{IwaPar}
V (q^\ta, A_\alpha) = \exp\left(q^\ta H_\ta\right) \exp\left(\sum_{\alpha> 0} \Aof{\alpha} \Eof{\alpha}\right).
\end{align}
The worldline model describing the motion of a point particle on the coset manifold
$G/K$ is then parametrised by a map $V: \reals \rightarrow G/K$, where
$t\in\reals$ is the time coordinate. The Cartan derivative is (with $\partial \equiv d/dt$)
\begin{align}
\label{CF}
\partial V V^{-1} =  P+ Q = \partial q^\ta H_\ta + \sum_{\alpha>0} e^{q^\ta \alpha_\ta}D\Aof{\alpha} \Eof{\alpha}
\end{align}
where $Q\in\mathfrak{k}\, , P\in\mathfrak{k}^\perp$, and,
schematically and due to $\partial e^X e^{-X}= \partial X +\frac12 \lb X,\partial X\rb +\ldots$,
\begin{align}\label{DA}
DA_\alpha = \partial A_\alpha + \frac12 \sum_{\beta>0\atop \alpha-\beta>0} c_{\alpha-\beta,\beta}\Aof{\alpha-\beta} \partial \Aof{\beta} + \ldots.
\end{align}
Importantly, the Borel gauge implies a triangular expansion of $D \Aof{\alpha}$ where the
factors contributing to the terms quadratic in $\Aof{\gamma}$ on the r.h.s. are of {\em lower} height, whence the sum on the r.h.s. of (\ref{DA}) has only finitely many terms even for 
infinite-dimensional Kac--Moody algebras (a crucial fact for the calculability of the model).
The invariant Lagrangian is given by %($\partial\equiv \partial_t$)
\begin{align}
\label{Lag}
L = \frac12 \langle P | P  \rangle = \frac12 \partial q^\ta G_{\ta\tb}\partial q^\tb 
+ \sum_{\alpha>0} \Pof{\alpha} \Pof{\alpha},
\end{align}
where we have defined
\begin{align}
P = \partial q^\ta H_\ta +\sum_{\alpha} \Pof{\alpha} (\Eof{\alpha} + \Eof{-\alpha}),\quad\quad
\Pof{\alpha} = \frac12 e^{q^\ta\alpha_\ta} D\Aof{\alpha}.
\end{align}
The compact part is then given by
\begin{align}
Q = \sum_{\alpha>0} \Qof{\alpha} (\Eof{\alpha}-\Eof{-\alpha})
= \sum_{\alpha>0} \Pof{\alpha} (\Eof{\alpha}-\Eof{-\alpha})
\end{align}
where the equality $\Qof{\alpha} = \Pof{\alpha}$ is a consequence of the 
triangular gauge choice. 

The model has global $G$ symmetry and local $K$ symmetry that we use to fix the triangular gauge~\eqref{IwaPar} everywhere. The symmetries act by
\begin{align}
\label{symactions}
V(t) \to  k(t) V(t) g^{-1}\quad
\Longrightarrow\quad
P \to k P k^{-1},\quad
Q\to k Q k^{-1} + \partial k k^{-1}.
\end{align}
When the triangular gauge~\eqref{IwaPar} is fixed, a local compensating
$K$-transformation is required to restore the gauge for every $G$ transformation
that throws $V$ out of the triangular gauge.

The equations of motion of the coset model are
\begin{align}
\label{eomall}
D P \equiv \partial P - [Q,P] = 0.
\end{align}
(We note that this of course implies that the equations of the original coordinates $q^\ta$ and $\Aof{\alpha}$ are second order differential equations.)
For a given root component $\Pof{\alpha}$ this means
\begin{align}
\label{eomrt}
\partial \Pof{\alpha} =-\partial q^\ta \alpha_\ta \Pof{\alpha}+2\sum_{\beta>0} 
c_{\alpha+\beta,-\beta} \Pof{\beta}\Pof{\alpha+\beta}  \equiv
-\partial q^\ta \alpha_\ta \Pof{\alpha}+2\sum_{\beta>0} 
c_{\alpha,\beta} \Pof{\beta}\Pof{\alpha+\beta}
\end{align}
Note that (in contrast to (\ref{DA})) the sum on the r.h.s. contains terms of {\em ascending} height.

\subsection{Changes in the Kac--Moody case}

When the Lie algebra $\mathfrak{g}$ is an infinite-dimensional Kac--Moody 
algebra~\cite{Ka90}, the definition of the corresponding group $G$ requires 
more care, see for example~\cite{KP,Kumar}. Again we restrict to 
simply-laced algebras, and more specifically to symmetric generalised Cartan matrices 
with at most one line linking any two nodes. Of course, our primary interest here will
be with $\E$ and its maximal compact subgroup $K(\E)$.

There are now two types of roots of the algebra, called real and imaginary and they distinguished by their Cartan--Killing norm: Real roots $\alpha$ satisfy $\alpha^2=2$ and imaginary roots $\alpha^2\leq 0$. The generators corresponding to real roots are unique up to normalisation and can be denoted by $\Eof{\alpha}$ as above but the generators corresponding to imaginary roots can have non-trivial multiplicities and are more appropriately denoted by $\Eof{\alpha}^r$, where $r=1,\ldots,\mathrm{mult}(\alpha)$ labels an orthonormal basis (w.r.t. the Cartan--Killing metric) in the root space. We will write all generators in this way, keeping in mind that for real roots $r$ can only take one value. The commutation relations in the Cartan--Weyl basis (cf.~\eqref{CWBasisFD}) then have to account also for the multiplicities and become
\begin{subequations}
\label{CWBasisKM}
\begin{align}
\lb \Eof{\alpha}^r, \Eof{\beta}^s \rb &= \left\{ \begin{array}{cl} \sum\limits_{t=1}^{\mathrm{mult}(\alpha+\beta)}c_{\alpha,\beta}^{rst} \Eof{\alpha+\beta}^t &\textrm{if $\alpha+\beta\in \Delta$,}
\\[4mm]
\delta^{rs} \alpha^\ta H_\ta  & \textrm{if $\alpha=-\beta$,}\\[2mm]
0 & \textrm{otherwise,}
\end{array}\right.\\[2mm]
\lb H_\ta , \Eof{\alpha}^r \rb &= \alpha_\ta \Eof{\alpha}^r.
\end{align}
\end{subequations}
We note that we still have $c_{\alpha,\beta} = \pm 1$ if $\alpha, \beta$ and
$\alpha + \beta$ are all real, but this need no longer be true when any of these
roots is imaginary. The bilinear form (\ref{Norm1}) generalizes to
\begin{subequations}
\label{Norm2}
\begin{align}
\langle \Eof{\alpha}^r | \Eof{\beta}^s \rangle
&=\left\{\begin{array}{cl} \delta^{rs} &\textrm{if $\alpha=-\beta$},\\[1mm]
0 &\textrm{otherwise},
\end{array}\right.\\[2mm]
\langle H_\ta | H_\tb \rangle &= G_{\ta\tb}
\end{align}
\end{subequations}
where the metric $G_{\ta\tb}$ is now indefinite (and Lorentzian for hyperbolic Kac--Moody algebras).

The other important modification concerns the parametrisation of the elements 
of the formal coset space $G/K$ that, using the Iwasawa decomposition, could be 
given in the finite-dimensional case as in~\eqref{IwaPar}. Even at a purely formal level, 
and even if the sum in the exponent is truncated to a finite number of terms, it is not directly meaningful to parametrize a given element of the Kac--Moody group in the form
\begin{align}
V(q^\ta, A_\alpha^r)  \;\; ``\!=\!" \; \;\exp (q^\ta H_\ta) \,
\exp\left( \sum_{\alpha>0}\sum_{r=1}^{\mathrm{mult}(\alpha)} \Aof{\alpha}^r \Eof{\alpha}^r \right)
\end{align}
where  $\big\{ q^\ta , \Aof{\alpha}^r\big\} $ are local coordinates on the (infinite-dimensional) coset
manifold, one coordinate for each Lie algebra element $\Eof{\alpha}^r$. The reason is that 
the step operators $\Eof{\alpha}^r$ associated with imaginary roots are not (locally) nilpotent 
in any standard representation, and therefore the exponential is {\it a priori}
ill-defined.\footnote{The notion of local nilpotency is defined as follows: an operator $\Eof{\alpha}^r$ is locally nilpotent in a representation $V$ of $\mathfrak{g}$ if for all $x\in V$ there exists an $n_0 = n_0(x)$ such that
$$
\big(\Eof{\alpha}^r\big)^n (x) = 0 \;\; \mbox{for all $n> n_0$.}
$$
(In the adjoint representation, the action is simply by commutators: $ \Eof{\alpha}^r (x)=\mathrm{ad}\, \Eof{\alpha}^r (x) \equiv \left[\Eof{\alpha}^r,x\right]$.) Clearly this holds for any real root $\alpha$ in the adjoint or any standard representation, but is no longer true for imaginary roots. The intuitive picture for this statement in the adjoint representation is that all roots lie in a solid hyperboloid $\{\alpha^2\leq 2\}$ in a Lorentzian space. Imaginary roots point {\em into} the light-cone where infinitely many roots of $\mathfrak{g}$ lie whereas real roots point outside the light-cone and eventually will leave the solid hyperboloid.}
 For this reason, standard approaches to Kac--Moody groups involve writing down only exponentials of {\em real} root generators (that are nilpotent) and then defining the Kac--Moody group as the group generated by the products of these real root exponentials~\cite{KP}. 
Although such a treatment is mathematically well-defined, it does not solve by any 
means the problem of finding a manageable realisation of the Kac--Moody group, 
because different orderings of exponentials of a given set of real root generators 
will yield new group elements. Organizing these differently ordered exponentials 
is thus directly associated to the (unsolved) problem of classifying the independent 
elements of the associated root space (where the problem is to count and classify 
the inequivalent ways in which a given set of Chevalley generators can 
be `distributed' over a multi-commutator). In particular, a parametrization in
terms of fields associated only to real roots of $\E$, besides being incomplete, 
would also obscure the relation to the fields $\big\{ \Aof{\alpha}^r\big\}$, 
and therefore does not appear to lead to a convenient parametrisation of the 
coordinates on the coset space $G/K$.\footnote{But let us note that the highest weights
associated to the {\em gradient representations} of~\cite{Damour:2002cu} are, in fact, real roots.} 
%(Let us emphasize
%that the Iwasawa decomposition $G=KAN$ is nevertheless still applicable~\cite{DeMedts}.)

Irrespective of an explicit description of the coordinates on $G/K$ we can still generalise
the worldline $\sigma$-model to infinite-dimensonal cosets, and consider the
Cartan form
\begin{align}\label{VdV}
\partial V V^{-1} = P+Q = \partial q^\ta H_\ta \,+\, \sum_{\alpha>0} \sum_{r=1}^{\mathrm{mult}(\alpha)} \Pof{\alpha}^r \Eof{\alpha}^r
\end{align}
without spelling out the explicit parametrisation of $V$ and $\Pof{\alpha}^r$ in terms of coordinates and their time derivatives. The triangular structure on $N$ implies, however, that 
the $\Pof{\alpha}^r$ are all \textit{finite} combinations of coordinates and their 
derivatives, as we explained already after (\ref{VdV}).

The coset equations~\eqref{eomall} take the same form if the Lagrangian is the formal
extension of (\ref{Lag}) to the infinite-dimensional Kac--Moody algebra, using
the invariant bilinear form (\ref{Norm2}) on the Kac--Moody algebra. 
Therefore~\eqref{eomrt} becomes
\begin{align}
\label{eomKMrt}
\partial \Pof{\alpha}^t =-\partial{q}^\ta \alpha_\ta \Pof{\alpha}^t
\,+\, 2\sum_{\beta>0} \sum_{r,s} c_{\alpha+\beta,-\beta}^{r\,s\,t} \Pof{\beta}^r\Pof{\alpha+\beta}^s.
\end{align}
As we noted after (\ref{eomrt}) the r.h.s. is a sum over terms of {\em ascending} height,
and hence an infinite sum for infinite-dimensional $\mathfrak{g}$. This sum can be 
rendered finite and calculable only by consistently truncating the $\Pof{\alpha}^r$ to 
vanish beyond a given height, as is necessary for the comparison between supergravity 
and the $\E$ $\sigma$-model. More concretely, this can be done for example by choosing a grading on the root lattice and cutting off $\Pof{\alpha}^r$ after a certain degree~\cite{Damour:2004zy}.

\subsection{Canonical treatment}

We now analyse the canonical structure, by again considering  
the \textit{finite-dimensional} coset model~\eqref{Lag} first.
The canonical momenta from~\eqref{Lag} are
\begin{align}
\pi_\ta &= \frac{\partial L}{\partial \partial q^\ta} = G_{\ta\tb} \partial q^\tb,&\nn\\
\Piof{\alpha} &= \frac{\partial L}{\partial \partial \Aof{\alpha}} = \frac12 \left(e^{2 q^\ta\alpha_\ta}D\Aof{\alpha} +\frac12 \sum_{\beta>0} c_{\beta,\alpha} e^{2 q^\ta(\alpha+\beta)_\ta}\Aof{\beta}D\Aof{\alpha+\beta} +\ldots\right),
\end{align}
displaying again a triangular structure. This can be inverted to write the $\Pof{\alpha}$ in triangular form in terms of canonical coordinates and momenta:
\begin{align}
\label{pi2p}
\Pof{\alpha} = e^{- q^\ta\alpha_\ta}\left( \Piof{\alpha} -\frac12 
\sum_{\beta>0} c_{\beta,\alpha}\Aof{\beta} \Piof{\alpha+\beta}+\ldots\right).
\end{align}
{}From this and the standard relations
\begin{equation}
\big\{ q^\ta, \pi_\tb \big\} = \delta^\ta_\tb \;, \quad
\big\{ \Aof{\alpha} , \Piof{\beta} \big\} = \delta_{\alpha,\beta}
\end{equation}
one can derive the canonical brackets among the $\pi^\ta$ and $\Pof{\alpha}$:
\begin{subequations}
\label{Pcan}
\begin{align}
\label{pipi}
\big\{ \pi_\ta, \pi_\tb\big\} &=0,&\\[2mm]
\label{piP}
\big\{ \pi_\ta, \Pof{\alpha} \big\} &= \alpha_\ta \Pof{\alpha},&\\[2mm]
\label{PP}
\big\{ \Pof{\alpha},\Pof{\beta} \big\} &=  c_{\alpha,\beta} \Pof{\alpha+\beta}.&
\end{align}
\end{subequations}
Only the first two of these relations are evident, while the third one is not and will be proven below.
Of course, to the order given one can check the last relation easily from the expressions above,
but the important point is that all the higher non-linear terms combine in the right way
to produce such a simple result. 
Our main point here is that the `composite' variables  $\Pof{\alpha}$ are `good'  
canonical variables because the canonical brackets between them assume a very simple form,
and furthermore display a graded structure which is nothing but the Borel subalgebra.
Equally important, the $\Pof{\alpha}$, being objects associated with the maximal
compact subgroup $K$, couple naturally to the fermions. Let us note the 
relations %(CHECK!)
\begin{equation}
\ld \pi_\ta , V \rd = - H_\ta V\; , \quad
\ld \Pof{\alpha} , V \rd = -   \Eof{\alpha} V \,,\quad
\ld \Pof{\alpha}, q^\ta \rd =0\,,\quad
\ld V, V \rd =0.
\end{equation}
For any coset space $\sigma$ model the canonical conserved Noether current 
(or more properly, conserved {\em charge}) is given by general formula
\begin{equation}
\label{Jcur}
J = V^{-1} P V \equiv  J^\ta  H_\ta \, + \, 
\sum_{\alpha > 0} \big( J_{-\alpha} \Eof{\alpha} + J_{\alpha} \Eof{-\alpha} \big)
\end{equation}
such that $\partial J = 0$ by the equations of motion (\ref{eomrt}).
Although the canonical commutation relations for this current reproduce the $GL(n)$
algebra (see below), we will see that the structure of its components is considerably more complicated,  not least because $J$ has both upper {\em and} lower triangular pieces.
For finite-dimensional $\mathfrak{g}$ the lower triangular half of the matrix $J$ takes 
a relatively simple form when one expresses the associated conserved components 
in terms of the momenta $\Piof{\alpha}$. By contrast the components of the 
upper triangular half involve all canonical variables and become increasingly 
more complicated with growing $n$, see also~\cite{Damour:2002et}. We will illustrate this explicitly with the example of the $GL(3)/SO(3)$ model in Appendix~\ref{app:GL3}.

The  conserved current $J$ of~\eqref{Jcur} generates the global $G$ transformations in~\eqref{symactions} and, since we are working in fixed triangular gauge, the `lower triangular' $G$ transformations induce a compensating $K$ transformation. That is, we expect the infinitesimal transformation of $V$ to be
\begin{align}
\ld J, V \rd = \delta{V} = - V \delta{g} + \delta{k} V,
\end{align}
where $\delta{g}$ and $\delta{k}$ are the infinitesimal versions of the group transformations in~\eqref{symactions} and $\delta{k}$ is determined by $\delta{g}$ and $V$ such that the resulting $\delta{V}$ is in triangular gauge. This can be worked out in terms of the basis components in~\eqref{Jcur} and the canonical brackets (\ref{Pcan}), with 
the result (for $\alpha >0$)
\begin{subequations}
\label{JV}
\begin{align}
\big\{ \Jof{\alpha} \,,\, V \big\} &= - V E_\alpha       \\[2mm]
\big\{ J_\ta \,,\, V \big\} &= - V H_\ta       \\[2mm]
\big\{ \Jof{-\alpha} \,,\, V \big\} &= - VE_{-\alpha}  - 
\sum_{\beta>0} \big\langle VE_{-\alpha}V^{-1} \big| E_\beta \big\rangle \,
    (E_\beta - E_{-\beta}) V,
\end{align}
\end{subequations}
where the extra term on the r.h.s. in the last line corresponds to the compensating
transformation in $K(G)$ required to bring $V$ back into triangular gauge, so that
\begin{equation}
\big\langle V^{-1} \{\Jof{-\alpha} , V \} \big| E_\beta \big\rangle = 0
\end{equation}
and the explicit compensating element in $\mathfrak{k}$ is
\begin{align}
\delta{k}_\alpha = -\sum_{\beta>0} \big\langle VE_{-\alpha}V^{-1} \big| E_\beta \big\rangle \,
    (E_\beta - E_{-\beta}).
\end{align}
In deriving the above brackets we made repeated use of the invariance of the 
invariant bilinear form (trace) and the orthonormality relations (\ref{Norm1}).\footnote{The invariance relation $\langle V^{-1}AV|B\rangle = \langle A|VBV^{-1}\rangle$ for $A,B\in
  {\mathfrak{g}}$ and $V\in G$ are also valid for infinite-dimensional $\mathfrak{g}$. In the finite-dimensional case, the relation can be viewed as the standard cyclic property of the matrix trace.}
Relation~\eqref{symactions} then implies directly the transformation of the velocity components $P$ under $J$. For $\alpha>0$ and $\beta>0$ one has
\begin{subequations}
\begin{align}
\ld J_\ta, \Pof{\beta} \rd &=0,\\
\ld \Jof{\alpha}, \Pof{\beta} \rd &=0 \\
\ld \Jof{-\alpha}, \Pof{\beta} \rd %&= \lb \delta{k}_\alpha,\Pof{\beta}\rb \nn\\
&=\pi_\ta \beta^\ta \langle V \Eof{-\alpha} V^{-1} |\Eof{\beta}\rangle -\sum_{\gamma>0} c_{\beta-\gamma,\gamma} \Pof{\gamma} \langle V \Eof{-\alpha} V^{-1} |\Eof{\beta-\gamma}\rangle\nn\\
&\quad\quad -\sum_{\gamma>0} c_{\beta+\gamma,-\gamma} \Pof{\gamma} \langle V \Eof{-\alpha} V^{-1} |\Eof{\beta+\gamma}\rangle 
\end{align}
\end{subequations}

Because (\ref{JV}) equivalently expresses the standard non-linear realisation
of global symmetries in non-linear $\sigma$-models  we can immediately infer 
the closure relations 
\begin{subequations}
\label{Jalgebra1}
\begin{align}
%\label{Jalgebra1}
\big\{ \Jof{\alpha}, \Jof{\beta} \big\} 
&= \left\{ \begin{array}{cl}  c_{\alpha,\beta} \Jof{\alpha+\beta} &\textrm{if $\alpha+\beta\in \Delta$,}
\\[2mm]
\alpha^\ta J_\ta  & \textrm{if $\alpha=-\beta$,}\\[2mm]
0 & \textrm{otherwise,}
\end{array}\right.\\[2mm]
\big\{ J_\ta , \Jof{\alpha} \big\} &=  \alpha_\ta \Jof{\alpha}.
\end{align}
\end{subequations}  
in the non-linear realization of the $G$ symmetry acting on $V$. An explicit
verification of the above relations for the $GL(3)/SO(3)$ model can be found
in appendix~\ref{app:GL3}.

An important aspect of the variables $\Pof{\alpha}$ concerns quantisation.
When quantising a non-linear model there is always the question for which canonical
variables one should perform the replacement of Poisson or Dirac brackets by quantum commutators
(which in quantum field theory may yield {\em inequivalent} quantisations). Obviously,
the variables $\Pof{\alpha}$ are ideally suited for this purpose; in particular,
such a quantisation prescription eliminates all operator ordering ambiguities.
Furthermore, we emphasise once again that the $\Pof{\alpha}$ are the 
natural variables coupling to fermions, as will be seen in more detail below.

\subsection{Canonical structure for $GL(n,\mathbb{R})/SO(n)$}
\label{sec:gln}

We now prove~\eqref{Pcan}, and in particular the crucial third relation, for  $G= GL(n,\reals)$. 
This is a slight generalisation of the set-up of the preceding sections since $GL(n,\reals)$ is not simple,
but it is the case of direct interest for cosmological billiards~\cite{Damour:2002et}. 

Let us fix some notation. We denote the generators of $GL(n,\reals)$ by $K^a{}_b$ with $a,b=1,\ldots,n$ and commutation relations
\begin{align}
\lb K^a{}_b, K^c{}_d\rb = \delta^c_b K^a{}_d - \delta^a_d K^c{}_b.
\end{align}
The symmetric and antisymmetric combinations are defined as $S^{ab} = K^a{}_b+K^b{}_a$ and $J^{ab} = K^a{}_b- K^b{}_a$. The positive roots of $GL(n,\reals)$ are denoted by $\alpha_{ab}$ with $a<b$ and will be written as tuples $\alpha_{ab} = (0\cdots 010\cdots 0 -\!\!1 0...)$ with $(+1)$ in the $a$-th and $(-1)$ in the $b$-th place. The generator corresponding to such a positive $\alpha_{ab}$ 
is then $\Eof{\alpha_{ab}}= K^a{}_b$ and the above commutation relations translate into 
(recall that $a<b$ and $c<d$)
\begin{equation}
\lb \Eof{\alpha_{ab}} , \Eof{\alpha_{cd}} \rb 
\,= \, c_{\alpha_{ab},\alpha_{cd}} \Eof{\alpha_{ab}+\alpha_{cd}} \, = \,
\left\{ \begin{array}{cl} \Eof{\alpha_{ad}} &\textrm{if $b=c$,}\\[1mm]
                          - \Eof{\alpha_{cb}} & \textrm{if $a=d$,}\\[1mm]
0 & \textrm{otherwise.}
\end{array}\right.
\end{equation}
So we read off the general formula $c_{\alpha_{ab},\alpha_{cd}}= \delta_{bc}-\delta_{ad}$.

We write the coset element of $GL(n,\reals)/SO(n)$ in Borel gauge by an 
upper triangular $(n\times n$)-matrix through (as for notation, cf. footnote 1)
\begin{align}
V=A N \quad\textrm{with} \quad A=\mathrm{diag}(e^{q^1},\ldots,e^{q^n})\;,
\quad N=N^a{}_i.
\end{align}
Here, $a$ is the local (row) index, $i$ a global (column) index. The matrix $N^{a}{}_i$ 
is equal to $1$ on the diagonal and has vanishing entries for $a>i$. The inverse matrix $N^{-1}$
is also upper triangular, and has components $N^i{}_a$ which vanish for $i>a$. The Borel gauge implies that some of the summations below are restricted in the way indicated. For the coset velocities one finds
\begin{align}
\left(\partial V V^{-1}\right)^a{}_b =   \partial q^a \delta^a_b + 
\sum_i e^{q^a- q^b}\partial N^a{}_i N^i{}_b.
\end{align}
For a positive root $\alpha_{ab}$, i.e. $a<b$, we define the quantity
\begin{align}
\Pof{\alpha_{ab}} = \frac12 e^{q^a- q^b} \sum_{a<i\leq  b} \partial N^a{}_i N^i{}_b
\end{align}
This is just the variable $\Pof{\alpha}$ introduced in the previous section, except that
we are now labelling the roots by indices $a,b$. In a convenient normalisation
the Lagrangian can be written as
\begin{align}
\label{LGL}
L \, = & \, \frac12 \mathrm{Tr}(P^2) - \frac12(\mathrm{Tr} P)^2 \nn\\[2mm]
   =  & \, \frac12 \partial q^\ta \partial q^\tb G_{\ta\tb} +  \sum_{a<b} \Pof{\alpha_{ab}} \Pof{\alpha_{ab}},
\end{align}
where $G_{\ta\tb}$ is the DeWitt metric
\begin{align}
\sum_{\ta,\tb}\partial q^\ta \partial q^\tb G_{\ta\tb} = \sum_\ta (\partial q^\ta)^2 -
 \left(\sum_\ta \partial q^\ta\right)^2.
\end{align}

The canonical momenta conjugate to $N^a{}_i$ are
\begin{align}
\Pi^i{}_a = \frac{\partial L}{\partial \partial  N^a{}_i} = 
\sum_b e^{q^a- q^b} \Pof{\alpha_{ab}} N^i{}_b
\end{align}
and vanish for $a\geq i$. In other words, for $a<b$,
\begin{align}
\label{Pab}
\Pof{\alpha_{ab}}=  \sum_i e^{q^b-q^a} N^a{}_i \Pi^i{}_b.
\end{align}

The advantage of using the variables $\Pof{\alpha}$ is that %, although they are complicated functions of the coset variables, 
they obey very simple commutation relations, {\it to wit},
\begin{align}
\big\{ \Pof{\alpha_{ab}}, \Pof{\alpha_{cd}}\big\} = c_{\alpha_{ab},\alpha_{cd}} \Pof{\alpha_{ab}+\alpha_{cd}}
\end{align}
whenever $\alpha_{ab}+\alpha_{cd}$ is a root, and we recall $c_{\alpha_{ab},\alpha_{cd}} =\delta_{bc}-\delta_{ad}$. This can be verified by straightforward computation using
$\{ N^a{}_i , \Pi^j{}_b \} = \delta^a_b \delta^j_i$. It is equally easy to see that if
$\alpha_{ab}+\alpha_{cd}$ is not a root, the canonical bracket vanishes, as does the structure constant. We thus have demonstrated the relation~\eqref{PP} for $GL(n,\reals)$.
%\begin{align}
%\label{PaPb}
%\big\{ \Pof{\alpha} , \Pof{\beta} \big\} =c_{\alpha,\beta} \Pof{\alpha + \beta}.
%\end{align}
The other relations~\eqref{pipi} and~\eqref{piP} are also evident for $GL(n,\reals)$,
given the explicit form of the Lagrangian~\eqref{LGL} and~\eqref{Pab}. 

The proof of (\ref{PP}) can be extended to all other simple finite dimensional
Lie algebras, either by direct computation, or more simply by looking at differently
embedded $GL(n)$ subalgebras, and by observing that these commutation relations
must be compatible with the action of the Weyl group, because all roots can be reached by 
Weyl transformations from the simple roots.

As a simple example we discuss the case of $GL(3)/SO(3)$ in  appendix~\ref{app:GL3}.

The extension of the above results to {\em infinite-dimensional} Kac--Moody algebras, 
and more specifically to the $\E$ algebra, is more subtle, and here we do not have a
complete picture. In particular, we do not have a proof that (\ref{piP}) and (\ref{PP})
remain valid for all roots. For instance, in the presence of imaginary roots (\ref{PP}) 
would have to generalise to
\begin{equation}
\label{PaPb1}
\big\{ \Pof{\alpha}^r , \Pof{\beta}^s \big\} =  \sum_{t=1}^{{\rm mult}(\alpha)} c^{rst}_{\alpha,\beta} \Pof{\alpha + \beta}^t.
\end{equation}
where $\alpha$ and $\beta$ are any roots, and where the sum on $t$ ranges over 
the multiplicity of the root $(\alpha + \beta)$ if this is an imaginary root. While a
general derivation by the above methods seems beyond reach, we can extend 
the argument at least to those roots $\alpha$ and $\beta$ for which $\alpha + \beta$
is also a real root, because the above commutation relation should respect the Weyl group,
and because all real roots can be reached by $\E$ Weyl transformations. Hence 
at least for this special case, the above relation should also hold for $\E$. 
%A further
%generalisation concerns the bracket between $\Pof{\alpha}^r$ and $\Pof{\beta}^s$
%when $\alpha$ and $\beta$ are imaginary roots, with $r$ and $s$ as extra labels
%for take into account the fact that their multiplicities are generally $> 1$

\subsection{Hamiltonian analysis}

The canonical Hamiltonian is
\begin{align}
H &= \pi_\ta\partial q^\ta +\sum_{\alpha>0} \Piof{\alpha} \partial \Aof{\alpha} - L&\nn\\
&= \frac12 \pi_\ta G^{\ta\tb} \pi_\tb + \sum_{\alpha>0} e^{-2q^\ta\alpha_\ta} \Piof{\alpha}^2 + \ldots
\end{align}
where the dots denote important non-linear terms. In terms of the coset velocities they can be summarised as (see also~\cite{Matschull:1994vi} for a derivation in the finite-dimensional case)
\begin{align}
\label{HFD}
H =\frac12  \pi_\ta G^{\ta\tb} \pi_\tb+ \sum_{\alpha>0} \Pof{\alpha}^2.
\end{align}
Again, it is important that the non-linear terms combine in the right way to yield
such a simple expression.
We note that~\eqref{Jcur} implies that we can rewrite the Hamiltonian alternatively as
\begin{align}\label{H=JJ}
H= \frac12 \langle J | J\rangle.
\end{align}
which we recognise as the standard bilinear form on the corresponding Lie algebra.

Let us verify the consistency expression~\eqref{HFD} with the coset equations of motion
\begin{align}
\partial \Pof{\alpha} = \left\{ \Pof{\alpha}, H \right\} = -\alpha_\ta G^{\ta\tb}\pi_\tb \Pof{\alpha} +2\sum_{\beta>0} c_{\alpha,\beta} \Pof{\beta}  \Pof{\alpha+\beta}.
\end{align}
%(Note that the cocycle properties of the structure constants imply that $c_{\alpha,\beta} = c_{\alpha+\beta,-\beta}$.)
Comparing the above relation with the general result~\eqref{eomrt} shows agreement. This shows that the Borel structure is correct, at least for all finite-dimensional algebras: any other 
algebra would not correctly reproduce the equations of motion.

Staying at the formal level, an analogous argument also works for the 
infinite-dimensional case. Namely, we can similarly deduce a statement of the 
canonical brackets of the $\Pof{\alpha}^r$ in the Kac--Moody case. Starting from 
the same Lagrangian
\begin{align}
L = \frac12 \langle P | P \rangle
\end{align}
as in the finite-dimensional case, but where $\langle\cdot|\cdot\rangle$ is now 
the standard invariant bilinear form on the Kac--Moody algebra, the  Hamiltonian is
given by the straight-forward formal extension of~\eqref{HFD}, using the arguments of~\cite{Matschull:1994vi}:
\begin{align}
H=\frac12  \pi_\ta G^{\ta\tb} \pi_\tb+ \sum_{\alpha>0} \sum_r \Pof{\alpha}^r \Pof{\alpha}^r
\end{align}
Because, formally, the conserved Noether current is still given by
\begin{equation}
\label{JcurKM}
J = V^{-1} P V \equiv  J^\ta  H_\ta \, + \, 
\sum_{\alpha > 0} \sum_{r=1}^{{\rm mult}(\alpha)} \big( J_{-\alpha}^r \Eof{\alpha}^r 
+ J_{\alpha}^r \Eof{-\alpha}^r \big)
\end{equation}
the Hamiltonian can again be cast into the form (\ref{H=JJ}) with the 
bilinear form on the Kac--Moody algebra. When considered as a function of  the
phase space variables $\big\{ J^\ta , J_{\pm \alpha}^r \big\}$
this is just the (unique) $\E$ invariant bilinear form. Let us mention, however,
that in contrast to the finite-dimensional case a simple form of the lower triangular half
can only be achieved by truncating the current components to 
$J_{-\alpha}^r =0$ for $\alpha$'s exceeding a given height. A related discussion can be found in~\cite{Damour:2002et}.

Compatibility of the canonical structure with the equations of motion~\eqref{eomKMrt} is then ensured by the canonical brackets %(CHECK multiplicity index order)
\begin{align}\label{PcanE10}
\ld \Pof{\alpha}^r , \Pof{\beta}^s \rd =\sum_t c^{r\,s\,t}_{\alpha,\beta} \Pof{\alpha + \beta}^t.
\end{align}
We thus see that {\em if} the Hamiltonian is given by the restriction of the $\E$
Casimir operator to the coset $\E/K(\E)$, the compatibility of the canonical structure
with the equations of motion {\em implies} the extension of the Borel-like structure found
in (\ref{Pcan}) to the full Borel subalgebra of $\E$! However, it is known that beyond
level $\ell = 3$ the canonical supergravity Hamiltonian starts to deviate from the Casimir
operator, and therefore we will also have to eventually allow for modifications in the canonical
algebra (\ref{PcanE10}).

\section{Fermions and supersymmetry}
\label{sec:fermions}

The extension of the $\E$ coset model to include fermions was discussed in~\cite{Damour:2005zs,deBuyl:2005mt,Damour:2006xu}. We briefly review the salient features of the resulting model and its relation to maximal $D=11$ supergravity in order to provide a self-contained presentation.

\subsection{$\E$, its level decomposition and the bosonic sector}

The description of $\E$ that is most commonly used in connection with $D=11$ supergravity is that where the Lie algebra is presented in $GL(10)$ level decomposition~\cite{Damour:2002cu}. In this presentation, the infinitely many generators of $\E$ are organised into $\mathfrak{gl}(10,\reals)$ tensor representations and graded by a level $\ell$ such that each level only contains finitely many $\mathfrak{gl}(10,\reals)$ representations. The Lie bracket is compatible with the level. At low non-negative levels one finds the following $\mathfrak{gl}(10,\reals)$ representations corresponding
to the (spatial) components of the $D=11$ fields and their magnetic duals:

\vspace{5mm}
\begin{center}
\begin{tabular}{c|c|l}
Level $\ell$ & Generator & Representation of $\mathfrak{gl}(10,\reals)$\\
\hline
\hline
$0$ & $K^a{}_b$ & $\mathbf{100}$ (adjoint; graviton)\\[2mm]
$1$ & $E^{abc}=E^{[abc]}$ & $\mathbf{120}$ (three-form)\\[2mm]
$2$ & $E^{a_1\ldots a_6}= E^{[a_1\ldots a_6]}$ & $\mathbf{210}$ (six-form)\\[2mm]
$3$ & $E^{a_0|a_1\ldots a_8}= E^{a_0|[a_1\ldots a_8]}$ with $E^{[a_0|a_1\ldots a_8]}=0$ & $\mathbf{440}$ ((8,1)-hook; dual graviton)
\end{tabular}
\end{center}
\vspace{5mm}

The `coset velocity' $P$ of~\eqref{CF} can be similarly decomposed by level
\begin{eqnarray}
P &=& \sum_{\alpha>0} \sum_{r=1}^{{\rm mult} (\alpha)}\Pof{\alpha}^r (\Eof{\alpha}^r + \Eof{-\alpha}^r ) 
\equiv \sum_{\ell\geq 0} P^{(\ell)} * E^{(\ell)}   \nn\\[2mm]
&\equiv& \frac12 P^{(0)}_{ab} S^{ab} 
 + \frac1{3!}  P^{(1)}_{abc} S^{abc} 
 + \frac1{6!} P^{(2)}_{abcdef} S^{abcdef} 
  + \frac1{9!} P^{(3)}_{a_0| a_1\cdots a_8} S^{a_0|a_1\cdots a_8} + \cdots
\end{eqnarray}
Here, the $P^{(\ell)}$ transform in the representation from the table branched to $SO(10)$ level (since $P$ transforms covariantly under the `compact' subgroup $K(\E)$). The generators are defined by
\begin{align}
S^{ab} &= K^a{}_b + K^b{}_a,&
S^{abc} &= E^{abc} + F_{abc},\nn\\[2mm]
S^{a_1\ldots a_6} &= E^{a_1\ldots a_6} + F_{a_1\ldots a_6},&
S^{a_0|a_1\ldots a_8} &= E^{a_0|a_1\ldots a_8} + F_{a_0|a_1\ldots a_8},
\end{align}
where $F_{abc}$ etc. are the Chevalley transposed generators on the negative levels and correspond to the $\Eof{-\alpha}^r$ part in the general expression. 

As was shown in~\cite{Damour:2002cu,Damour:2006xu}, the bosonic coset model with Lagrangian $L=\frac12 \langle P|P \rangle$, when restricted to levels $\ell\leq 3$,  
is equivalent to $D=11$ supergravity expanded about a fixed spatial point, 
$\xnull$ with the bosonic dictionary\footnote{\label{convchange}We have here adjusted some normalisations relative to~\cite{Damour:2006xu} in order to make subsequent expressions more uniform. The changes concern the dual fields on levels $\ell=2$ and $\ell=3$: The sign on level two here is opposite to that of~\cite{Damour:2006xu}, and $P^{(3)}_{\textrm{here}} = \frac13 P^{(3)}_{\textrm{DKN}}$. The reason for the rescaling is that we here are normalising the real root generators identically on all levels in conformity with (\ref{Norm2}).}
\begin{align}
\label{dictbos}
P^{(0)}_{ab} (t) &= -N \omega_{ab 0} (t,\xnull),&
P^{(1)}_{abc}(t) &= N F_{0abc}(t,\xnull),\\[2mm]
P^{(2)}_{a_1\ldots a_6}(t) &= \frac1{4!} N \epsilon_{a_1\ldots a_6 b_1\ldots b_4} F^{b_1\ldots b_4}(t,\xnull),&
P^{(3)}_{a_0|a_1\ldots a_8}(t) &= \frac12 N \epsilon_{a_1\ldots a_8 b_1 b_2} \omega_{b_1 b_2 a_0}(t,\xnull)
\end{align}
when all higher order spatial gradients are neglected and the $SO(10)$ connection is traceless $\omega_{bba}=0$ (corresponding to the irreducibility condition of the $\ell=3$ generator in the table) and $t$ is the coordinate along the worldline that is identified with the physical time coordinate. The index $0$ is a flat index in the time-direction and $N$ is the lapse function in ADM gauge with zero shift. With the said truncations it can then be shown that the bosonic equations
of motion of $D=11$ supergravity coincide with those of the worldline $\E$ sigma model.

In order to re-express these $SO(10)$ objects in terms of $\E$ root data and $\Pof{\alpha}^r$ we need to explain how the roots at the various levels are related to the components. As in section~\ref{sec:Can}, we work in the so-called `wall basis'~\cite{Damour:2009zc,Damour:2013eua}. This means that we write a root $\alpha$ as $\alpha=\sum_\ta \alpha_\ta e^\ta$ where $e^\ta$ are the basis of the $\mathfrak{h}^*$ dual to the Cartan generators $H_\ta$ such that 
$e^\ta(H_\tb) = \delta^\ta_\tb$ and hence $\alpha(H_\ta) = \alpha_\ta$. 
In the wall basis, the inner product is given by 
\begin{equation}\label{eaeb}
\langle e^\ta| e^\tb \rangle = G^{\ta\tb} = \delta^{\ta\tb}-\frac19
\end{equation} 
and agrees with the (inverse) DeWitt metric for diagonal metrics.  In order to avoid confusion 
with the labelling of the simple roots it will sometimes be convenient to also use the notation
$ p_\ta \equiv \alpha_\ta$ interchangeably for the component of $\alpha$ in the wall 
basis, and also write $\alpha=(p_1,\ldots, p_{10})$ as a row vector. 
The ten simple roots of $\E$ are explicitly given by
\begin{align}
\alpha_1 &= (1,-1,0,0,\ldots,0),\nn\\
\alpha_2 &= (0,1,-1,0,\ldots,0),\nn\\
\vdots&\nn\\
\alpha_9 &= (0,\ldots,0,0, 1,-1),\nn\\
\alpha_{10} &= (0,0,\ldots,0,1,1,1).
\end{align}
The $\mathfrak{gl}(10)$ level of an arbitrary root $\alpha$ expanded on the simple roots as 
$\alpha= \sum_{j=1}^{10} m^j \alpha_j$ is $\ell\equiv\ell(\alpha) = m_{10}$.

Roots on level $\ell=0$ are roots of $\mathfrak{gl}(10)$ and can be written as $\alpha_{ab}$ as above in section~\ref{sec:gln}. The components $\Pof{\alpha}$ for these roots are identified with $P_{ab}^{(0)}$ and we let $a<b$ for positive roots as before. The components of the Cartan subalgebra are identified via $\pi^\ta = P^{(0)}_{aa}$ (no sum).
Roots on level $\ell=1$ have three entries $1$ in the wall basis and the other entries $p_\ta$ are zero, as for example in $\alpha_{10}$ above. Calling the three non-vanishing components $a$, $b$ and $c$ with $a<b<c$, we identify $\Pof{\alpha}$ with $P_{abc}^{(1)}$. Roots $\alpha_{a_1\ldots a_6}$ on level $\ell=2$ have six entries $p_\ta=1$ and four vanishing entries. We assume $a_1<\ldots<a_6$ and then identify $\Pof{\alpha_{a_1\ldots a_6}}$ with the corresponding level $\ell=2$ coset velocity. Roots on level $\ell=3$ come in two varieties: They either have one entry $p_\ta=2$, seven entries $p_\ta=1$ and two $p_\ta=0$ or they have nine $p_\ta=1$ and one $p_\ta=0$. In the first case, we let $a_0$ be the component with entry $p_{a_0}=0$ and assume again that the $p_\ta=1$ components are ordered as $a_1<\ldots a_7$. Then we identify $\Pof{\alpha}$ with $P_{a_0|a_0a_1\ldots a_7}^{(3)}$. The second case corresponds to null roots (of multiplicity $8$) and schematically we distribute the ordered nine $p_\ta=1$ components as $\Pof{a_0|a_1\ldots a_8}$. The multiplicity requires extra care and will be discussed in detail in section~\ref{sec:gauge}. 

In summary, we find that we can associate
\begin{align}
\label{newbosons}
\Pof{\alpha_{ab}} = P^{(0)}_{ab},\quad
\Pof{\alpha_{abc}} = P^{(1)}_{abc},\quad
\Pof{\alpha_{a_1\ldots a_6}} = P^{(2)}_{a_1\ldots a_6},\quad
\Pof{\alpha_{a_0|a_1\ldots a_8}} = P^{(3)}_{a_0|a_1\ldots a_8}.
\end{align}
with root labels on the l.h.s., and with the associated $SO(10)$ tensors on
the r.h.s. Up to $\ell\leq 3$, this correspondence rule allows us to rewrite any 
expression involving $P^{(\ell)}$ in terms of $\Pof{\alpha}$.

\subsection{Unfaithful spinor representations of $K(\E)$}
\label{sec:KE10}

Fermions are associated with the compact subalgebra $K(\E)$ of $\E$. 
This algebra is generated by the compact combinations ($\alpha>0$)
\begin{align}
\label{Jalpha}
\kof{\alpha}^r = \Eof{\alpha}^r - \Eof{-\alpha}^r.
\end{align}
We have chosen the 
letter $\kof{\alpha}^r$ for the $K(\E)$ generators, rather than using $J(\alpha)^r$ 
as in~\cite{Kleinschmidt:2013eka} in order to avoid confusion with the components of the conserved current $J$ in~\eqref{Jcur}. From~\eqref{CWBasisKM} the $K(\E)$ elements satisfy
\begin{align}\label{kcommutator}
\big[  \kof{\alpha}^r\,,\, \kof{\beta}^s \big]  = 
\sum_{t=1}^{{\rm mult}(\alpha + \beta)} c_{\alpha,\beta}^{rst} \,\kof{\alpha +\beta}^t \;-\;
\sum_{t=1}^{{\rm mult}(\alpha - \beta)} c_{\alpha,-\beta}^{rst} \kof{\alpha-\beta}^t.
\end{align}
In order to make sense of the above relation in general, and because $\alpha-\beta$ can be $<0$ for $\alpha,\beta>0$, one also requires a 
definition of  $\kof{\alpha}^r$ for $\alpha<0$; from (\ref{Jalpha}) we directly get
\begin{equation}
\label{Jminus}
\kof{\alpha}^r := - \kof{-\alpha}^r \qquad \mbox{ for $\alpha < 0$}.
\end{equation}
which is also consistent with (\ref{cab}).

$K(\E)$ admits unfaithful finite-dimensional spinor representations~\cite{Damour:2005zs,deBuyl:2005mt,Damour:2006xu,deBuyl:2005zy,Koehl1}, but unfortunately no faithful spinor representations are
known up to now. The unfaithful representations relevant to supergravity involve the vector-spinor (gravitino) and Dirac-spinor (supersymmetry parameter). The representations can be represented conveniently using the wall basis~\cite{Damour:2009zc,Damour:2013eua} and we use the same formalism as in~\cite{Kleinschmidt:2013eka}. For the Dirac representation it is enough
to restrict attention to real roots $\alpha,\beta,\dots$, and we will thus drop the 
multiplicity labels in the remainder of this section. Then to every element $v$ of the
$\E$ root lattice $v = \sum n^j \alpha_j \equiv \sum_\ta v_\ta e^\ta$  (which need not 
be a root for arbitrary $n^j \in \ints$)
we associate an element of the $SO(10)$ Clifford algebra through
\begin{align}\label{Gamma}
\Gamma(v)= (\Gamma_1)^{v_1} \cdots (\Gamma_{10})^{v_{10}},
\end{align}
where, of course, $\{ \Ga_a , \Ga_b\} = 2\delta_{ab}$ are the usual $SO(10)$ $\Ga$-matrices.
The product of two such matrices is given by
\begin{equation}\label{Guv}
\Ga(u) \, \Ga(v) = \veps_{u,v} \Ga(u \pm v),
\end{equation}
where we have defined the cocyle
\begin{align}
\veps_{u,v} = (-1)^{\sum_{\ta<\tb} v_\ta u_\tb}.
\end{align}
which obeys
\begin{equation}\label{cocycle}
\veps_{u,v} \veps_{v,u} = (-1)^{u\cdot v} \quad , \qquad
\veps_{u,v} \veps_{u+v,w} = \veps_{u,v+w} \veps_{v,w}
\end{equation}
where $v\cdot w \equiv G^{\ta\tb} v_\ta w_\tb$.
The cocycle $\veps_{u,v}$ is defined only up to a co-boundary, that is,
we can modify the above definition (\ref{Gamma}) by
\begin{equation}\label{Gamma1}
\Ga (v) \quad \rightarrow \quad  \Gt(v) = \si_v \Ga(v)
\end{equation}
with $\si_v = \pm 1$ an (in principle) arbitrary sign factor; then
\begin{equation}
\vepst_{u,v} = \si_u \si_v \si_{u+v} \veps_{u,v}
\end{equation}
also obeys the cocycle relations (\ref{cocycle}). Next we specialize to elements 
$v= \alpha\,,\, \beta \in \Delta$ which {\em are} roots, and choose the co-boundary such that
\begin{equation}\label{Gamma2}
\Gt(\alpha) := \sigma_\alpha \Gamma(\alpha) :=
       \left\{\begin{array}{cl} \Ga(\alpha) &\textrm{if $\alpha > 0$},\\[2mm]
             - \Ga(\alpha)\equiv -\Gamma(-\alpha) &\textrm{if $\alpha < 0$},
\end{array}\right.
\end{equation}
that is, $\si_\alpha = \pm 1$ according to whether $\alpha$ is positive or negative,
whence $\si_\alpha \si_{-\alpha} = -1$. The sign switch between positive and negative 
roots in (\ref{Gamma2}) is necessary to remain consistent with (\ref{Jminus}).
This definition can be extended to the whole root lattice by choosing 
$\si_v = \pm 1$ arbitrarily for non-roots $v$, but subject to the 
condition $\si_v \si_{-v} = -1$ (for $v\neq 0$).
Indeed, in the relevant expressions in the supersymmetry constraint the
matrix $\Gt(\alpha)$ always comes with a factor $\Pof{\alpha}^r$ which vanishes
when $\alpha$ is not a root. The extra sign in (\ref{Gamma2}) leads to an important
modification in the multiplication rule (\ref{Guv}), {\it viz.}
\begin{equation}\label{Gamma3}
\Gt(\alpha) \, \Gt(\beta) \,=\, \vepst_{\alpha,\beta} \Gt(\alpha + \beta)
    \,=\,  - \vepst_{\alpha,-\beta} \Gt(\alpha - \beta).
\end{equation}
With these definitions one can check that the map 
\begin{equation}\label{Diracrep}
\kof{\alpha}\; \mapsto \; \frac12\, \Gt(\alpha)
\end{equation}
for all real $\alpha$ provides a representation of $K(\E)$, when extended consistently by commutators. For consistency of this representation with (\ref{Jminus}) the sign 
in~\eqref{Gamma2} is crucial. 
This representation has a large kernel;  for instance, for null roots $\delta$ 
one has $\kof{\delta}^r\mapsto 0$, and for time-like imaginary roots all elements
of the corresponding root space either vanish or are represented by the {\em same} element of the
Clifford algebra. The quotient algebra of $K(\E)$ by the kernel is isomorphic to $\mathfrak{so}(32)$~\cite{Damour:2006xu}. We will refer to this representation of $K(\E)$ as the Dirac-spinor representation, or just `Dirac representation'. (This type of representation can be straight-forwardly generalised to other simply-laced Kac--Moody algebras and also to arbitrary Kac--Moody algebras~\cite{Koehl1}.) 

Because (\ref{Diracrep}) works for all real roots the comparison of  (\ref{kcommutator}) with 
(\ref{Gamma3}) shows that, for real roots $\alpha$ and $\beta$,
\begin{equation}
c_{\alpha,\beta} = -\vepst_{\alpha,\beta}
\end{equation}
whenever $\alpha+ \beta$ or $\alpha-\beta$ is also a real root. This is furthermore consistent 
with the fact that for real $\alpha$ and $\beta$ only one of the terms on the r.h.s. of
(\ref{kcommutator}) can be non-zero. The minus sign in the above relation arises because below we will act on the components of the spinor rather than on the basis vectors. 

While the Dirac representation corresponds to the supersymmetry transformation
parameter, the vector spinor representation derives from the $D=11$ gravitino, and
is `less unfaithful' than the Dirac representation. It was first obtained in~\cite{Damour:2006xu}
in terms of an $SO(10)$ covariant vector-spinor $\Psi^a_A$ with an $SO(10)$ vector index 
$a=1,\ldots,10$ and spinor indices $A,B,\ldots= 1,\ldots,32$.  This vector-spinor is 
directly related via a fermionic dictionary to the spatial components of the $D=11$ gravitino $\psi_a$ through~\cite[Eq.~(5.1)]{Damour:2006xu}:
\begin{equation}
\label{dictferm}
\Psi_a(t) = g^{1/4} \psi_a(t,\xnull) ,
\end{equation}
where $g=\det (g_{mn})$ is the determinant of the spatial part of the metric. For the time
component of the gravitino (the Lagrange multiplier for the supersymmetry
constraint), we adopt the gauge $\psi_0 = \Ga_0 \Ga^a \psi_a$, 
as in \cite{Damour:2006xu}.

For the present purposes it is, however, advantageous to switch to a different
description of the vector-spinor in terms of fermions $\phi^\ta$ %for~\eqref{phispinors} 
which are related to the $SO(10)$ covariant vector-spinor $\Psi^a$ of~\cite{Damour:2006xu} 
above by the following crucial re-definition~\cite{Damour:2009zc}
\begin{align}
\label{newfermions}
\phi^\ta = \Gamma^a \Psi^a \quad\quad \textrm{(no sum on $a$)}.
\end{align}
This relation clearly breaks $SO(10)$ covariance, but has an important advantage:
in this way the Lorentz group $SO(10)$ gets replaced by the $SO(1,9)$ symmetry acting
on the space of diagonal scale factors $\{ q^\ta\}$, which is also the invariance
group of the DeWitt metric $G_{\ta\tb}$!  It is for this reason that we adopt a
different font ($\ta, \tb,...$), as we already did in \cite{Kleinschmidt:2013eka}; the
latter indices are then covariant under the (Lorentzian) invariance group of
the DeWitt metric $G_{\ta\tb}$. We will also use the notation 
\begin{align}
\phi(\alpha) \equiv  \alpha_\ta \phi^\ta.
\end{align}
Like the Dirac representation the vector-spinor representation, now modeled by 
spinors $\phi^\ta_A$, is obviously finite-dimensional (we will often suppress explicitly 
writing out the spinor indices). The vector-spinor $\phi^\ta_A$ satisfies the 
canonical (Dirac) brackets~\cite{Damour:2006xu,Damour:2009zc}:
\begin{align}
\label{phispinors}
\ld \Psi^a_A , \Psi^b_B \rd = \delta^{ab}\delta_{AB} - \frac19 (\Ga^a \Ga^b)_{AB} \;\; 
\Rightarrow \quad  \ld \phi^\ta_A, \phi^\tb_B \rd = G^{\ta\tb} \delta_{AB}.
\end{align}
(recall the definition of $G^{\ta\tb}$ in (\ref{eaeb})).
A canonical representation of $K(\E)$ is then obtained by
defining for any real root $\alpha$
\begin{align}\label{Jreal}
\kof{\alpha} = X_{\ta\tb}(\alpha)\phi^\ta \Gt(\alpha) \phi^\tb, \quad
X_{\ta\tb}\equiv X_{\ta\tb}(\alpha) = - \frac12 \alpha_\ta \alpha_\tb + \frac14 G_{\ta\tb}.
\end{align}
and this construction yields an unfaithful representation 
of $K(\E)$~\cite{Kleinschmidt:2013eka}. Note that we again have to employ the 
$\Gt$-matrices from (\ref{Gamma2}) in order to extend this definition to both positive
and negative real roots. We also note that the unfaithful spinor representation can be used to deduce partial information about the unknown structure constants of $K(\E)$, and thus $\E$.

With the bosonic dictionary~\eqref{dictbos} and the fermionic dictionary~\eqref{dictferm} one can now convert any supergravity expression into the $\E$-variables $P^{(\ell)}$ and $\Psi_a$. With the relations~\eqref{newbosons} and~\eqref{newfermions} we can then rewrite in the next step everything into $\Pof{\alpha}$ and $\phi^\ta$ variables. This is the procedure we now apply to the supersymmetry constraint of $D=11$ supergravity.

\subsection{Supersymmetry constraint}

In terms of the original canonical variables of $D=11$ supergravity~\cite{Cremmer:1978km}, the canonical 
supersymmetry constraint is given by~\cite[Eq. (3.12)]{Damour:2006xu}
\begin{eqnarray}
\label{cSt}
\tilde{\cS} &=& \Gamma^{ab} \Big[ \partial_a \psi_b + \frac14 \omega_{acd} \Gamma^{cd} \psi_b 
     + \omega_{abc} \psi_c + \frac12 \omega_{ac0} \Gamma^c \Gamma^0 \psi_b \Big] \nn\\[1mm]
     && +  \, \frac14 \, F_{0abc} \Gamma^0 \Gamma^{ab} \psi^c + \frac1{48} F_{abcd} 
     \Gamma^{abcde} \psi_e,
\end{eqnarray}
where $\omega_{ABC}$ are the components of the $D=11$ spin connection and 
$F_{ABCD}$ the components of the four-form (with flat indices $A,B,...=0,1,\dots,10$).
Using the dictionaries~\eqref{dictbos} and~\eqref{dictferm} one can rewrite this expression in terms of $\E$ coset variables. The translation between the coset model and $D=11$ supergravity furthermore involves neglecting spatial gradients $\partial_a \psi_b$ on the fermions, terms of the form $\partial_a g\propto \omega_{bba}$, and all spatial gradients of second or higher order on the bosonic fields.
It was then shown in~\cite[Eq. (5.14)]{Damour:2006xu} that the supersymmetry constraint 
can be re-expressed in terms of the coset quantities $P^{(\ell)}$ and in an $SO(10)$ covariant
manner as
\begin{align}
\label{SUSYconstr}
 \cS = & \left( P^{(0)}_{ab}\Ga^a - P^{(0)}_{cc} \Ga_b\right)\Psi^b 
       + \, \frac12 P^{(1)}_{abc} \Ga^{ab} \Psi^c
       + \, \frac1{ 5!} P^{(2)}_{abcdef} \Ga^{abcde} \Psi^f  \nn\\[2mm]
  & + \, \frac1{ 6!} \left(P^{(3)}_{a|ac_1\cdots c_7} \Ga^{c_1\cdots c_6} \Psi^{c_7}
      - \frac1{28} P^{(3)}_{a|c_1\cdots c_8} \Ga^{c_1\cdots c_8} \Psi^a \right) . 
\end{align}
Compared to~\cite{Damour:2006xu}, we have rescaled the supersymmetry constraint by an overall factor of $2$ and we also recall the normalisation changes that we explained in footnote~\ref{convchange}.

The notation $\cS$ in place of $\tilde{\cS}$ of~\eqref{cSt} indicates that we have rescaled $\tilde{\cS}$ and multiplied it by $\Gamma_0$. 
%The fermionic variables $\Psi_a$ appearing in~\eqref{SUSYconstr} are the $K(\E)$ vector-spinor fields and related to the supergravity variables in~\eqref{cSt} by $\Psi_ a = g^{1/4} \psi_a$.
In this $SO(10)$ covariant form, repeated indices are summed over and indices are raised and lowered with the Euclidean metric $\delta_{ab}$. We will now rewrite this expression once more, in order to bring it into a form that conforms more closely with the new variables
introduced in the foregoing section. A key fact here is that by so doing we will give up manifest
spatial Lorentz covariance, and trade it for the Lorentzian $SO(1,9)$ symmetry on the 
space of scale factors exhibited above. In other words, the simplest form of the constraint is
attained by trading a space-time symmetry for a symmetry in (a truncated version of)
DeWitt superspace!

To convert the expression~\eqref{SUSYconstr} to the $\E$ covariant notation above, we change fermionic variables according to~\eqref{newfermions} and analyse the various terms.
For the contributions from $\ell=0,1,2$, and now writing out the sums,  we find 
\begin{eqnarray}
\sum_a P^{(0)}_{aa} \Ga^a\Psi^a  - \sum_c P^{(0)}_{cc} \sum_a \Ga_a \Psi^a 
    &=&   %G_{ab} \pi^a \phi^b   \, \equiv \, 
    G_{\ta\tb} \pi^\ta \phi^\tb        \nn\\[1mm]
 \sum_{a<b}  P^{(0)}_{ab}\Ga^a \Psi^b \, + \, \sum_{a>b}  P^{(0)}_{ab}\Ga^a \Psi^b 
    &=&  \sum_{a<b} P^{(0)}_{ab} \Ga^{ab} (\phi^b - \phi^a) \nn\\[1mm]
\sum_{a,b,c} P^{(1)}_{abc} \Ga^{ab} \Psi^c &=& 
%\frac13 \sum_{a,b,c} P^{(1)}_{abc} \Ga^{abc} (\phi^a + \phi^b + \phi^c)  \nn\\[1mm]
2 \sum_{a<b<c} P_{abc}^{(1)} \Gamma^{abc}(\phi^a + \phi^b + \phi^c)  \nn\\[1mm]
 \sum_{a,b,c,d,e,f} P^{(2)}_{abcdef} \Ga^{abcde} \Psi^f &=&
% \frac16 \sum_{a,b,c,d,e,f} P^{(2)}_{abcdef} \Ga^{abcdef}  (\phi^a + \cdots + \phi^f)  
5! \sum_{a<b<c<d<e<f} P^{(2)}_{abcdef} \Ga^{abcdef}  (\phi^a + \cdots + \phi^f)  
\end{eqnarray}
where we identified $\pi^\ta= P^{(0)}_{aa}$. We now see that the expressions on the r.h.s.
are already in the desired form; for instance,
\begin{align}
\label{sumneg}
  \sum_{a<b<c} P_{abc}^{(1)} \Gamma^{abc}(\phi^a + \phi^b + \phi^c) \, &\equiv \,
  \sum_{\alpha_{abc}} P_{\alpha_{abc}} \Ga(\alpha_{abc}) \phi(\alpha_{abc}) 
\nonumber\\[2mm]  
   &=  \, \frac12 \sum_{\ell(\alpha) = \pm 1} \Pof{\alpha} \, \Gt(\alpha) \phi(\alpha)    
\end{align}
where the middle sum on the r.h.s. runs over all level $\ell=1$ roots $\alpha_{abc}$ (which
are positive), while the last sum includes positive and negative roots. The level $\ell =0 , 2$
contributions work in an analogous manner.

%Recalling the explicit expressions for the $\E$ roots at levels $\ell=0,1,2$ from~\eqref{},
%and reverting to indices $\ta,\tb,\dots$ the sum of the above terms can be written in the form [SIGN on $\ell=2$??]
%\begin{equation}
%\Ga^0 \cS_A  \Big|_{\ell\leq 2} \, = \, G_{\ta\tb} \pi^\ta \phi^\tb_A + 
%\sum_{\ell(\alpha) = 0,1,2} \Pof{\alpha} \Gamma(\alpha)_{AB} \,\phi(\alpha)_B.
%\end{equation}
%This expression exhibits a nice uniform structure in terms of the roots.

At level $\ell=3$ we encounter not only real roots, but for the first time also null roots.
To see this distinction one has to separately analyse those terms in
$P^{(3)}_{a|c_1\cdots c_8}$ for which
the index $a$ coincides with one of the $c_i$ (yielding real roots), and those terms
for which all indices are different, i.e. $a\notin \{ c_1, \dots , c_8 \}$ (yielding null roots).
In order to analyse these terms we thus have to split up the various sums. We start with
\begin{align}
 \sum_{a,c_1,\ldots,c_7} P^{(3)}_{a|ac_1\cdots c_7} \Ga^{c_1\cdots c_6} \Psi^{c_7} &\equiv 
 \sum_{c_1,\ldots, c_7} \sum_{a\neq c_i}  P^{(3)}_{a|ac_1\cdots c_7} \Ga^{[c_1\cdots c_6} \Psi^{c_7]}\nn\\[2mm]
 &= 6!  \sum_{c_1<\cdots<c_7}\sum_{a\neq c_i}  P^{(3)}_{a|ac_1\cdots c_7}  \Ga^{c_1\cdots c_7} (\phi^{c_1}+\ldots +\phi^{c_7}),
\end{align}
where the $c_i$ have been ordered in the second expression. The other contribution to the supersymmetry constraint~\eqref{SUSYconstr} becomes
\begin{align}
&-\frac1{28}\sum_{a} \sum_{c_1,\ldots,c_8} P^{(3)}_{a|c_1\cdots c_8} \Ga^{c_1\cdots c_8} \Psi^a 
=-2\cdot 6! \sum_{a}\sum_{c_1<\cdots<c_8} P^{(3)}_{a|c_1\cdots c_8} \Ga^{c_1\cdots c_8} \Psi^a \nn\\[2mm]
&\quad\quad\quad= 2\cdot 6!\sum_{c_1<\cdots<c_7} \sum_{a\neq c_i} P^{(3)}_{a|ac_1\cdots c_7} \Ga^{c_1\cdots c_7} \phi^a 
-2\cdot 6! \sum_{c_1<\cdots<c_8}\sum_{a\neq c_i} P^{(3)}_{a|c_1\cdots c_8} \Ga^{ac_1\cdots c_8} \phi^a. 
\end{align}
Combining the two parts one finds
\begin{align}
& \hspace{30mm}\frac1{6!} \left(P^{(3)}_{a|ac_1\cdots c_7} \Ga^{c_1\cdots c_6} \Psi^{c_7}
      - \frac1{28} P^{(3)}_{a|c_1\cdots c_8} \Ga^{c_1\cdots c_8} \Psi^a \right)  \nn\\[2mm]
&=  \sum_{c_1<\cdots<c_7} \sum_{a\neq c_i} P^{(3)}_{a|ac_1\cdots c_7} \Ga^{c_1\cdots c_7} (2\phi^a + \phi^{c_1}+\ldots +\phi^{c_7})
- 2 \, \sum_{c_1<\cdots<c_8}\sum_{a\neq c_i} P^{(3)}_{a|c_1\cdots c_8} \Ga^{ac_1\cdots c_8} \phi^a. 
\end{align}
The first term is exactly the contribution from the $360$ (gravitational) real roots 
on level $\ell=3$, {\it viz.}
\begin{equation}
\alpha = (2111111100) \quad \mbox{and permutations}
\end{equation}
where the root shown is associated with the component $P^{(3)}_{1|12345678}$.
The normalization is different from the one used previously
since the level $\ell=3$ generators were normalised to $9$ rather than $1$ 
in~\cite{Damour:2006xu}, cf. also footnote~\ref{convchange}. The second term is a sum over the (gravitational) null roots 
 \begin{equation}
 \label{nullrts}
\delta = (1111111110) \quad \mbox{and permutations}
\end{equation}
where the first root is now associated with the component $P^{(3)}_{1|23456789}$.
Note that the constraint as written above  is overcounting them 
since there are $\begin{pmatrix}10\\8\end{pmatrix}\times 2 = 90$ instead of the required $80$. The reason is a new type of gauge invariance related to the irreducibility of the $\ell=3$ representation and that will be discussed in more detail in section~\ref{sec:gauge}.

Let us summarise: altogether, the rewriting of the supersymmetry 
constraint~\eqref{SUSYconstr} so far has led to the following expression up to and
including all roots of $\ell\leq 3$:
\begin{align}
\label{SUSY3}
 %\Gamma^0
 \mathcal{S} &= \pi\cdot\phi + \sum_{\alpha^2=2\atop{\ell=0}, \alpha>0} \Pof{\alpha} \Gt(\alpha) \phi(\alpha)
+ \sum_{\alpha^2=2\atop{\ell=1}} \Pof{\alpha}\Gt(\alpha) \phi(\alpha)
+ \sum_{\alpha^2=2\atop{\ell=2}} \Pof{\alpha} \Gt(\alpha) \phi(\alpha)\nn\\
&\quad+  \sum_{\alpha^2=2\atop{\ell=3}} \Pof{\alpha} \Gt(\alpha) \phi(\alpha)
+ \sum_{\delta^2=0\atop{\ell=3}} \sum_{r=1}^8 \Pof{\delta}^r \Gt(\delta) \phi(\epsilon^r)
\end{align}
where we have replaced $\Ga$ by $\Gt$ to underline that the sum can also be extended to run over
negative roots as in~\eqref{sumneg}. The `polarisation vectors' $\epsilon^r$ appearing for the null roots will be discussed in detail in the next section.
%(INCLUDE FACTORS 1/2????).
%The signs and factors could be eliminated by changes of conventions but we adhere to the ones here for easier comparison with the previous literature.

\subsection{Null roots and gauge equivalences}
\label{sec:gauge}

We now return to the counting issue mentioned after~\eqref{nullrts}.  The association of a particular index set $(a_1 c_1\ldots c_8)$ with all indices different to a null root component  $P_{a_1|c_1\ldots c_8}^{(3)}$ is subject to the irreducibility constraint (Young symmetry)
\begin{align}\
\label{irred}
P_{[a_1|c_1\ldots c_8]}^{(3)} =0.
\end{align}
This provides one linear relation between a priori nine different ways of distributing the nine indices on the hook tableau, bringing down the number of independent components to eight, in agreement with the multiplicity of null roots in $\E$. Let us discuss in more detail how this is implemented in the supersymmetry constraint.

{\bf Gauge fixed form:}
To see this in more detail let us pick the particular null root corresponding to the indices $\{a,c_1,\ldots,c_8\}=\{1,\ldots, 9\}$. This root has contributions proportional to $\Gamma(\delta)$ through (reordering some of the indices)
\begin{align}
-2\left(P_{1|2\ldots 9} \phi^1 + P_{2|3\ldots 9 1} \phi^2 + \ldots +P_{9|1\ldots 8} \phi^9\right)
\end{align}
Here, one could now trade the first term for a combination of the other terms by virtue of~\eqref{irred}. This leads to 
\begin{align}
-2\left(P_{2|3\ldots 9 1} (\phi^2-\phi^1) + \ldots +P_{9|1\ldots 8} (\phi^9-\phi^1)\right),
\end{align}
that is, it can be written in the form
\begin{align}
\sum_{r=1}^8 \Pof{\delta}^r \Gamma(\delta) \phi(\epsilon^r)
\end{align}
with 
\begin{equation}
\Pof{\delta}^1 = P_{2|3\ldots 9 1}\;, \,\ldots,\; \Pof{\delta}^8=P_{9|1\ldots8 }
\end{equation}
and polarization vectors
\begin{equation}
\epsilon^1 = (2\,-\!2\, 0\, 0\, 0 \; 0\, 0\, 0\, 0\, 0),\,\ldots,\, \epsilon^8 = (2\, 0\, 0\, 0\, 0 \; 0\, 0\, 0\, -\!2\, 0)
\end{equation}
These polarisation vectors are orthogonal to $\delta$ (as required) and correspond 
to positive $\ell=0$ roots associated with generators $K^1{}_{r+1}$ (or their negatives).

{\bf Gauge unfixed form}:
 We can avoid choosing a particular set of polarisation vectors by instead letting the `multiplicity sum' run over an enlarged set
\begin{equation}
\sum_{r=1}^9\Pof{\delta}^r\Gamma(\delta) \phi(\epsilon^r).
\end{equation}
Here, $\Pof{\delta}^r$ denote the nine index arrangements and $\epsilon^r$ are nine independent polarisation vectors that are orthogonal to $\delta$. Shifting $\epsilon^r\to \epsilon^r+\delta$ leads to
\begin{align}
\sum_{r=1}^9 \Pof{\delta}^r\Gamma(\delta) \phi(\epsilon^r+\delta)= \sum_{r=1}^9 \Pof{\delta}^r\Gamma(\delta) \phi(\epsilon^r)+ \sum_{r=1}^9 \Pof{\delta}^r\Gamma(\delta) \phi(\delta) =\sum_{r=1}^9 \Pof{\delta}^r \Gamma(\delta) \phi(\epsilon^r)
\end{align}
since $\sum_{r=1}^9 \Pof{\delta}^r=0$ by virtue of~\eqref{irred}. Therefore, we have a 
gauge invariance in the expression that we could use to fix the gauge in the way we 
have done above. This gauge invariance no longer `lives' in ordinary space-time, but
rather in the DeWitt superspace of (diagonal) metrics.

\section{Properties of supersymmetry constraint}

Having rewritten the supersymmetry constraint in terms of $K(\E)$ variables
we will now re-investigate the canonical algebra of supersymmetry constraints
and its $K(\E)$ covariance. As for the algebra we will recover the previously
derived results according to which the canonical constraints of $D=11$ supergravity
in the appropriate truncation are all associated with null roots of $\E$. As for the 
transformation properties of the superconstraint, we will exhibit its non-covariance
under the full $K(\E)$ -- a clear indication that the present construction is incomplete.

\subsection{Supersymmetry constraint  algebra}

The above calculations led to the following expression for the supersymmetry constraint
\begin{align}\label{S}
\cS_A &= \pi_\ta \phi^\ta_A + 
\sum_{\alpha^2=2\atop{\ell\leq 3},\alpha>0} \Pof{\alpha} \big(\Gamma(\alpha) \phi(\alpha)\big)_A + 
\sum_{\delta^2=0\atop{\ell =3}} \Pof{\delta}^r \big(\Gamma(\delta)\phi(\epsilon^r)\big)_A
%&\nn\\
%&\qquad\qquad + \sum_{\alpha^2} J(\alpha) \Gamma(\alpha)\phi(\alpha) + \ldots
\end{align}
(recall that  $\phi(\alpha)_A \equiv \alpha_\ta \phi^\ta_A$).
%and where the dots stand for terms beyond $\ell=3$ (and the truncated supergravity expression) that we will discuss in more detail in section~\ref{sec:beyond}.
%Note that the terms written out contain {\em all} roots up to $\ell\leq 3$, where the level $\ell$ refers to the decomposition of $\E$ under its $GL(10)$ subgroup (obtained by removing the exceptional node from the Dynkin diagram). For $\ell\leq 2$ all roots are real; for $\ell =3$ there are both real and null roots, while beyond $\ell=3$ there will also appear imaginary time-like roots (with $\alpha^2 < 0$). 
As shown above, the terms written out in the above formula agree precisely 
with the supersymmetry constraint derived from supergravity by dropping terms
containing spatial gradients as well as cubic terms in the fermions.
In other words, apart from these omitted contributions,
the full content of the supersymmetry constraint is captured 
by the $\ell\leq 3$ sector of the $\E$ model with fermions. However, from this 
restriction it is already clear that this expression cannot be  the whole story, 
and we will make this point more explicit in the following section by showing
that, contrary to first expectations, $\cS$ does not transform in the Dirac representation,
nor in any other known representation of $K(\E)$.

Nevertheless, under the canonical brackets, the supersymmetry constraint in the above
form should yield the Hamiltonian and all other constraints supergravity constraints
in the gradient truncation (and ignoring higher order fermionic terms).
Schematically, we see that
\begin{align}\label{SS}
\big\{ \cS_A, \cS_B\big\} = 
2\cH \delta_{AB}  + \sum_{\delta^2=0}  \cC(\delta) \Gamma_{AB}(\delta) + \ldots
\end{align}
Here, we have introduced the calligraphic letter $\cH$ for the `Hamiltonian' arising from the commutator of two supersymmetry constraints, to distinguish it notationally from the 
coset Hamiltonian $H$ discussed in the previous sections, since it is not clear a priori 
whether the two agree. Indeed, we will explain below that they do differ.

The anti-commutator~\eqref{SS} contains many terms, but let us first concentrate on the ones 
containing no fermions (the ones bilinear in the fermions would also receive
contributions from cubic fermionic terms, which are not included in the above
formula for $\cS$). Here we use (for roots $\alpha$ and $\beta$ that are real and hence have anti-symmetric $\Gamma(\alpha)$ and $\Gamma(\beta)$)
\begin{subequations}
\begin{align}
\big\{ \pi\cdot\phi_A\, ,\,\pi\cdot \phi_B \big\} &= G_{\ta\tb} \pi^\ta\pi^\tb \delta_{AB} \\[2mm]
\big\{ \pi\cdot \phi_A\,,\, \Pof{\alpha} \big(\Gamma(\alpha)\phi(\alpha)\big)_B \big\} &=
   (\alpha \cdot \pi)  \Pof{\alpha} \Gamma(\alpha)_{BA} \, + \, \cdots \\[2mm]
\label{SSpart3}
\big\{ \pi\cdot \phi_A\,,\, \Pof{\delta}^r \big(\Gamma(\delta)\phi(\veps^r)\big)_B \big\} &=
   (\veps^r  \cdot \pi)  \Pof{\delta}^r \Gamma(\delta)_{BA} \, + \, \cdots \\[2mm]
\big\{ \Pof{\alpha} \big(\Gamma(\alpha)\phi(\alpha)\big)_A
 \, ,\, \Pof{\beta} \big(\Gamma(\beta)\phi(\beta)\big)_B \big\} &= 
   - (\alpha \cdot \beta)  \veps_{\alpha,\beta} \,
   \Pof{\alpha} \Pof{\beta} \Gamma(\alpha + \beta)_{AB} \, + \cdots
\end{align}
\end{subequations}
where dots stand for terms quadratic in the fermions.
Now the anticommutator ~\eqref{SS} is {\em symmetric} in $A,B$, hence the terms in the second
line do not contribute because $\Gamma(\alpha)$ is antisymmetric for real roots $\alpha$.\footnote{This is in agreement with structure of the diagonal components $\pi_\ta=P_{aa}^{(0)}$ from the constraint $C_{[a_1\ldots a_9]}^{(3)}= P_{ca_1}^{(0)} P_{c|a_2\ldots a_9}^{(3)}+\ldots$ discussed in~\cite{Damour:2007dt,Damour:2009ww} that only couple to the null root components $\Pof{\delta}^r$ as determined by~\eqref{SSpart3}.}
Consequently, the result will then contain only terms proportional to
$\delta_{AB}$ (the Hamiltonian), and terms where $\alpha + \beta$ is light-like 
(the constraints), and more generally, for which $(\alpha + \beta)^2$ is a multiple 
of four. This is indeed the structure displayed in (\ref{SS}).

Let us first look at the Hamiltonian. The first kind of contribution will come from those
terms with $\beta = \alpha$; in this case we use $\veps_{\alpha,\alpha} = -1$ to get
\begin{equation}\label{Ham1}
 - (\alpha \cdot\alpha)  \veps_{\alpha,\alpha} \,
   \Pof{\alpha} \Pof{\alpha} \, \Gamma(2\alpha)_{AB} 
   \, = \, + \, 2 \, \Pof{\alpha} \Pof{\alpha} \delta_{AB}
\end{equation}
which is positive, and agrees with what we get from the $\E$ Casimir (see below). For the second  
kind we have $\alpha\neq \beta$, but such that $(\alpha + \beta)$ has only even
components, such that again $\Gamma(\alpha + \beta) = {\bf 1}$; for example $\alpha+\beta=2\delta=(22222\,22220)$ with 
$$
\alpha = (21111\,11100) \quad \mbox{and} \quad \beta = (01111\,11120)
$$
In this case we still have $\veps_{\alpha,\beta} = -1$ but $\alpha\cdot\beta = -2$, hence
\begin{equation}\label{Ham2}
  - (\alpha \cdot \beta)  \veps_{\alpha,\beta} \,
   \Pof{\alpha} \Pof{\beta} \, \Gamma(\alpha + \beta)_{AB} 
   \, = \, - \, 2\,  \Pof{\alpha} \Pof{\beta} \delta_{AB}
\end{equation}
As one can easily check these are indeed associated with the {\em negative definite} 
terms in the bosonic part of the supergravity Hamiltonian. To see this more explicitly,
we recall from~\cite[Eq.~(6.6)]{Damour:2006xu} the $SO(10)$ covariant expressions 
for $\cH$ arising from the supersymmetry commutator:
\begin{align}\label{HSUGRA}
\cH &=
  \frac12 P^{(0)}_{ab}P^{(0)}_{ab}
  - \frac12 P^{(0)}_{aa}P^{(0)}_{bb}
  +\frac1{ 3!}P^{(1)}_{abc}P^{(1)}_{abc}
  +\frac1{ 6!}P^{(2)}_{a_1\ldots a_6}P^{(2)}_{a_1\ldots a_6}\nn\\[2mm]
&\quad+\frac2{8!} \Big( P^{(3)}_{a_0|a_1\ldots a_8}P^{(3)}_{a_0|a_1\ldots a_8}
  - 4  \, P^{(3)}_{b|ba_1\ldots a_7}P^{(3)}_{c|ca_1\ldots a_7} \Big)\nn\\[2mm]
  &= \frac12 \pi_\ta G^{\ta\tb} \pi_\tb \;+ \sum_{\alpha>0\atop \alpha^2=2,\ell\leq 3} \Pof{\alpha} \Pof{\alpha} \; - \sum_{\alpha,\beta>0,  \alpha+\beta=2\delta\atop \alpha^2=\beta^2=2,\ell= 3} \Pof{\alpha}\Pof{\beta}% \;+ \; \textrm{imaginary roots}
\end{align}
(see footnote~\ref{convchange} for the normalisations of the level-2 and level-3 terms). Writing out the sums in 
last two terms we get exactly the two contributions (\ref{Ham1}) and (\ref{Ham2})
(plus the contribution from the null root).
This result is to be contrasted with the coset Hamiltonian $H$
\begin{align}
\label{HCOS}
H= \frac12 \langle P | P\rangle &= \frac12 P^{(0)}_{ab}P^{(0)}_{ab}
  - \frac12 P^{(0)}_{aa}P^{(0)}_{bb}
  +\frac1{ 3!}P^{(1)}_{abc}P^{(1)}_{abc}
  +\frac1{ 6!}P^{(2)}_{a_1\ldots a_6}P^{(2)}_{a_1\ldots a_6}\nn\\[2mm]
& \quad  +\, \frac1{8!}P^{(3)}_{a_0|a_1\ldots a_8}P^{(3)}_{a_0|a_1\ldots a_8}
         \; + \, \ldots\nn\\[2mm]
&= \frac12 \pi_\ta G^{\ta\tb} \pi_\tb \;+ 
  \sum_{\alpha > 0\atop \alpha^2 =2,\ell\leq 3} \Pof{\alpha} \Pof{\alpha} %\; +\; \textrm{imaginary roots} 
      \; + \; \ldots
\end{align}
where the dots stand for higher level real roots, as well as imaginary roots.
In $SO(10)$ form, this latter expression differs from the previous one not 
only by the appearance of 
the negative $\ell =3$ term in (\ref{HSUGRA}) but also by the factor in front of the
$\ell=3$ term with the correct sign. However, when writing out the Hamiltonian (\ref{HSUGRA}) in terms of the $K(\E)$ variables
in a manner completely analogous to the derivation in the foregoing section, we see
that the terms at level $\ell=3$ are in one-to-one correspondence with the two types
of terms exhibited in (\ref{Ham1}) and (\ref{Ham2}).\footnote{We stress that  
the coefficients of the $\Pof{\alpha}\Pof{\alpha}$ terms for real roots do agree in 
the two expressions.  For $\ell=3$, this might seem surprising in view of the {\em different} 
coefficients in the $SO(10)$ covariant expressions. A simple way of seeing that they 
agree after the rewriting in $K(\E)$ variables is to look at a fixed particular real 
root, say $\Pof{\alpha_{1|12345678}}\equiv P^{(3)}_{1|12345678}$. 
In~\eqref{HSUGRA} this term has contributions from both expressions via 
$\frac{2}{8!}(8!-4\cdot 7!)=1$ and in~\eqref{HCOS} one similarly has $\frac1{8!}\cdot 8!=1$.
This example illustrates well how the $K(\E)$ properties can be obscured by
insisting on $SO(10)$ invariant expressions.}  We have thus isolated the 
source of the disagreement between the canonical Hamiltonian and the $\E$ Casimir
that appears from level $\ell =3$ onwards, in terms of the $K(\E)$ covariant looking
supersymmetry constraint (\ref{S}). This disagreement is seen not only in the
negativity, but also in the fact that the $\E$ Casimir does not pair $\Pof{\alpha}$
with $\Pof{\beta}$ for $\beta\neq \pm \alpha$. Clearly the source of the trouble resides
in the unfaithfulness of the $K(\E)$ representation in terms of the $\Gamma(\alpha)$ 
matrices that we are dealing with here, and would seem to require a generalisation
of the usual Clifford algebra. We also see that these troubles multiply when we extend 
the sum from roots with $\ell\leq 3$ to all real roots, as we will then have many more 
contributions proportional to $\delta_{AB}$, which would ruin the agreement with
the supergravity Hamiltonian found above.

The bosonic constraints identified in \cite{Damour:2006xu} and associated there with
lightlike roots are also recovered from those combinations where $\alpha + \beta$ is a
null root; as both $\alpha$ and $\beta$ can go up to level $\ell =3$, the resulting
null roots go up to level $\ell =6$, in agreement with \cite{Damour:2006xu}. So the
constraints are generically of the form
\begin{equation}
\cC(\delta) = \sum_r \veps^r \cdot  \Pof{\delta}^r \; + \,
\sum_{\alpha + \beta = \delta} \Pof{\alpha} \Pof{\beta} \, + \cdots
\end{equation}
which agrees exactly with what was found before in \cite{Damour:2006xu}. Note, however,
that starting from the supersymmetry constraint (\ref{S}), the first term on the r.h.s. 
only appears for the null root at level $\ell =3$, whereas this term is missing for
the higher level null roots, because the supersymmetry constraint only goes up to
$\ell =3$. By contrast, the null roots appearing in the combinations $\alpha + \beta$ can
go up to $\ell =6$. This is a clear signal of the incompleteness of the supersymmetry
constraint (\ref{S}) as derived from supergravity. We note also that there is only {\em one} constraint per null root $\delta$ whereas there are eight root generators $\Eof{\delta}^r$.

We note that the fermion $\phi^\ta$ appears as a {\em matter fermion} in the one-dimensional model even though it transforms in a vector-spinor representation and descends from the $D=11$ gravitino. This can for instance be seen by considering the transformation of $\phi^\ta$ under $\cS$ of~\eqref{SUSY3} which does not contain any derivatives of the transformation parameter (these would come from $D_a \phi^\ta$ terms that were truncated away in the derivation from supergravity). The one-dimensional gravitino that is the supersymmetry partner of the one-dimensional lapse function was set to zero.

%The Hamiltonian here is expected to be related to the coset part of the quadratic Casimir of $\E$
%\begin{align}
%\cH &= \pi^2 + \sum_{\alpha^2=2\atop{\ell\leq 3}} \Pof{\alpha}^2 + 
%\sum_{\delta^2=0\atop{\ell=3}} (\Pof{\delta}^r)^2 + \cdots
%\end{align}
%and as such is positive definite. As we will see in a moment, this result coincides with
%the one obtained from the supersymmetry algebra only up to $\ell =2$, but starts to
%deviate from $\ell =3 $ onward.
%The remaining terms are related to the canonical constraints
%\begin{align}
%\cC(\delta) &= \sum_{r}(\pi\cdot \epsilon^r) \Pof{\delta}^r +
%\sum_{\alpha^2, (\delta -\alpha)^2=2} \Pof{\alpha}\Pof{\delta-\alpha}  +\ldots
%\end{align}
%Here, the ellipsis indicates higher order terms in fermions and also additional contributions at the purely bosonic level.
%
%\begin{itemize}
%\item Work out all terms as far as possible
%\item Find a convenient Fierz identity that is needed for the $\phi^2$ and $\phi^4$ terms
%\item Think about (non-)$K(\E)$ covariance of $\cS$
%\end{itemize}

\subsection{(In)compatibility of supersymmetry and $K(\E)$}
\label{sec:susytrm}

We can now also investigate the transformation properties of the constraint $\cS$
under $K(\E)$. Because $\cS$ is `built'  out of objects that do transform properly
under $K(\E)$, namely the coset quantities $\Pof{\alpha}$ on the one hand,
and the unfaithful vector spinor representation $\phi^\ta$ on the other, one would
naively expect this constraint to transform in the Dirac representation, that is, 
$\delta_\alpha \cS = \frac12 \Gamma(\alpha) \cS$. However, there appears a basic 
clash: as we will now show very explicitly, $\cS$ {\em fails to transform properly under} $K(\E)$. 
There are two  reasons for this, namely first the presence of imaginary roots in $\E$ and
$K(\E)$ (and thus the fact that both algebras are infinite-dimensional), 
and secondly the unfaithfulness of the vector-spinor representation.
For the variation of $\cS$ under a $K(\E)$ transformation generated
by $\kof{\alpha}$ we use the formulas
\begin{eqnarray}
\delta_\alpha \pi^\ta &=&  -2\alpha^\ta \, \Pof{\alpha}  \nn\\[2mm]
\delta_\alpha \Pof{\beta} &=&  \delta_{\alpha,\beta} \alpha_\ta \pi^\ta \,  +\,
  c_{\beta-\alpha,\alpha} \Pof{\beta - \alpha} - c_{\alpha + \beta, -\alpha} \Pof{\alpha + \beta}
 \nn\\[2mm]
\delta_\alpha \phi^\ta &=& \frac12 \Gt(\alpha) \phi^\ta - \alpha^\ta \Gt(\alpha) \phi(\alpha)
\end{eqnarray}
restricting to positive real $\alpha, \beta$ for simplicity. For the first two lines we have evaluated $\lb P, \kof{\alpha}\rb$ and projected onto the $H_\ta$ and $\Eof{\beta}+\Eof{-\beta}$ components.\footnote{For other roots
the first two lines would generalize to 
\begin{eqnarray}
\delta^r_\alpha \pi^\ta &=& - 2\alpha^\ta \, \Pof{\alpha}^r  \nn\\[2mm]
\delta^r_\alpha \Pof{\beta}^s &=&  \delta^{rs} \delta_{\alpha,\beta} \alpha_\ta \pi^\ta \,  +\,
  \sum_t c_{\beta-\alpha,\alpha}^{rst} \Pof{\beta - \alpha}^t - 
  \sum_t c_{\alpha + \beta, -\alpha}^{rst}  \Pof{\alpha + \beta}^t     \nonumber
\end{eqnarray}
but no general formula is available for $\delta^r_\alpha \phi^\ta$. A conjectural formula is given in appendix~\ref{app:vsrep}.}
 We emphasize that it is not known whether the $\Pof{\alpha}$, when supplemented by the
higher root partners $\Pof{\alpha}^r$, transform in an irreducible representation of $K(\E)$, or whether this representation is reducible under $K(\E)$. Substituting these formulas into
the variation of $\cS$ some further calculation leads to
\begin{eqnarray}
\delta_\alpha \cS &=& \frac12 \Gt(\alpha) \cS \; + \;
\frac12 \sum_{\beta>0} \Pof{\beta} \big[ \Gt(\beta) , \Gt(\alpha)\big] \phi(\beta) \nn\\[2mm]
&& + \sum_{\beta>0\,,\, \beta\neq \alpha} \Big[ - (\alpha\cdot\beta) \Pof{\beta}
\Gt(\beta)\Gt(\alpha) \phi(\alpha) \, + 
\, c_{\beta-\alpha,\alpha} \Pof{\beta - \alpha} \Gt(\beta) \phi(\beta) \nn\\[2mm]
&& \qquad  \qquad \qquad\qquad\qquad
- \, c_{\alpha + \beta, -\alpha} \Pof{\alpha + \beta} \Gt(\beta) \phi(\beta) \Big]
\end{eqnarray}
The result would thus have the desired structure if we could show that all terms on 
the r.h.s. cancel except for the first. However, as we will now demonstrate by 
explicit computation this is the case  only for finite dimensional $K$, 
but no longer  for $K(\E)$. To do so we first rewrite the last term in brackets as
\begin{equation}
- \sum_{\beta>\alpha} c_{\beta,- \alpha} \Pof{\beta}\Gt(\beta - \alpha) \phi(\beta - \alpha)
\end{equation}
and shift the sums in the second term such a way that only $\Pof{\beta}$ with $\beta >0$ appear. 
So in the second term in brackets above we consider the partial sum\footnote{The inequalities 
in the sums are taken to imply that the corresponding elements are roots of the 
algebra (as an ordering cannot be generally defined for arbitrary pairs of
roots $\alpha$ and $\beta$).}
\begin{equation}
\sum_{0<\beta <\alpha} c_{\beta-\alpha,\alpha} \Pof{\beta- \alpha} \Gt(\beta) \phi(\beta) =
\sum_{0 < \beta < \alpha} c_{-\beta,\alpha} \Pof{-\beta} \Gt(- \beta+ \alpha) \phi(- \beta + \alpha)
\end{equation}
Next, using $\Pof{-\beta} = \Pof{\beta}$ and $\phi( - \beta + \alpha) = - \phi(\beta - \alpha)$
as well as  (\ref{cab}) and not forgetting the extra minus sign from the definition of $\Gt$ 
in (\ref{Gamma2}) for negative roots this term becomes equal to
\begin{equation}
- \sum_{0<\beta <\alpha} c_{\beta,- \alpha} \Gt(\beta-\alpha) \phi(\beta-\alpha)
\end{equation}
and therefore combines with the above term to give a full sum over $\beta >0$
(because $\phi(0) = 0$, there is no contribution for $\beta=\alpha$).  Finally we obtain
\begin{eqnarray}
\delta_\alpha \cS &=& \frac12 \Gt(\alpha) \cS \; + \;
\frac12 \sum_{\beta>0} \Pof{\beta} \big[ \Gt(\beta) , \Gt(\alpha)\big] \phi(\beta) \nn\\[2mm]
&& + \sum_{\beta>0\,,\, \beta\neq \alpha} \Big[ - (\alpha\cdot\beta) \Pof{\beta}
\Gt(\beta)\Gt(\alpha) \phi(\alpha) \, + 
\, c_{\beta,\alpha} \Pof{\beta} \Gt(\alpha + \beta) \phi(\alpha + \beta) \nn\\[2mm]
&& \qquad  \qquad \qquad\qquad\qquad
- \, c_{\beta, -\alpha} \Pof{\beta} \Gt(\beta - \alpha) \phi(\beta- \alpha) \Big]
\end{eqnarray}
where the sign in (\ref{Gamma2}) is again essential. Let us now inspect the 
different terms here: using $\phi(\alpha + \beta) = \phi(\alpha) + \phi(\beta)$
the terms containing $\phi(\alpha)$ become
\begin{equation}
-\sum_{\beta>0\, , \, \beta\neq \alpha}
\Pof{\beta}\Big\{ (\alpha\cdot\beta) \Gt(\beta)\Gt(\alpha)- c_{\beta,\alpha} \Gt(\alpha+\beta)
         - c_{\beta,-\alpha}\Gt(\beta-\alpha)\Big\}  \phi(\alpha)
\end{equation}
The expression inside brackets does indeed cancel 
if $\alpha$ and $\beta$ are real roots such that $(\alpha \pm \beta)$ are also 
real roots (in which case $\alpha\cdot\beta = \mp 1$); for $\alpha\cdot\beta =0$
all terms vanish. This covers all possible cases for $GL(n)$, but for indefinite
$G$ there are infinitely many more possibilities because $\alpha\cdot\beta$ can
assume any value, and then the extra terms no longer obviously cancel.  We note that there is some room for modifications of the argument coming from the values of $c_{\alpha,\beta}$ when $\alpha+\beta$ is imaginary and also from terms associated with imaginary roots in the ansatz~\eqref{SS}. The calculation in~\cite{Damour:2006xu} shows that the (truncated) expression~\eqref{SUSY3}, involving some terms from null roots, does not transform covariantly.

For the terms containing $\phi(\beta)$ the argument is similar; in this case we end up with
\begin{equation}
\frac12 \sum_{\beta >0 \, , \, \beta\ne\alpha} \Pof{\beta}\Big\{ \big[ \Gt(\beta) , \Gt(\alpha)\big] 
    + 2 c_{\beta,\alpha} \Gt(\alpha + \beta) - 2 c_{\beta,-\alpha} \Gt(\beta- \alpha)
    \Big\} \phi(\beta)
\end{equation}
and a case by case analysis analogous to the one above 
shows again that these terms cancel under the same conditions
as before. To sum up, the extra terms do cancel for finite dimensional $G$,
when we need only consider the cases $\alpha\cdot\beta = \pm1$ or $=0$;
in this case the supersymmetry constraint indeed transforms properly under $K$. 
This need no longer be true for infinite-dimensional $G$, where we have only 
insufficient knowledge of the structure constants $c_{\alpha,\beta}$.
Let us also emphasize that this problem arises already at linear order in the fermions, so
the addition of cubic or even higher order fermion terms cannot remedy this problem.

\section{Outlook}

In this section, we discuss various possible extensions of our results. One pressing challenge is the correct treatment of the full $\E$ algebra beyond level $\ell=3$ when trying to construct a $K(\E)$ covariant supersymmetry constraint. This will be discussed in sections~\ref{sec:beyond}. Irrespective of the construction of a supersymmetric model one can consider the spinning particle of~\cite{Damour:2006xu} and how the fermionic degrees of freedom influence the canonical structures discussed in section~\ref{sec:Can}. We offer some comments on this in section~\ref{sec:ferm} below.

\subsection{Tentative generalisation beyond $\ell=3$}
\label{sec:beyond}

The expression~\eqref{S} is very suggestive of a generalisation beyond level $\ell=3$,
so we are tempted to propose
\begin{align}\label{Smod}
\cS_A &= \pi_\ta \phi^\ta_A + 
\sum_{\alpha^2=2\atop{\ell\leq 3}} \Pof{\alpha} \big(\Gamma(\alpha) \phi(\alpha)\big)_A + 
\sum_{\delta^2=0\atop{\ell =3}} \Pof{\delta}^r \big(\Gamma(\delta)\phi(\epsilon^r)\big)_A
\, + \, \cdots
%&\qquad\qquad + \sum_{\alpha^2} J(\alpha) \Gamma(\alpha)\phi(\alpha) + \ldots
\end{align}
where the dots could stand for (at least) three kinds of additional terms, namely
\begin{enumerate}
\item additional terms linear in fermions associated with higher ($\ell>3$) level roots, coming either from real or imaginary roots; 
\item additional terms cubic in fermions;
\item terms involving new `higher spin' or other unfaithful realizations of $K(\E)$.
\end{enumerate}
In the following, we will concentrate only on the first extension. This already represents an extension beyond the truncated supergravity constraints. We note that the arguments of section~\ref{sec:susytrm} show that such a generalisation will not be $K(\E)$ covariant. Nevertheless we can find some constraints on the possible form by demanding at least Weyl invariance of the known terms.

% Let us first concentrate on the terms linear in $\phi^\ta_A$. These terms derive from
%SO(10) covariant expressions derived in section 5.3 of~\cite{Damour:2006xu}
%(hence we are still dealing with Lorentz indices $a,b,\dots$)
%We will now show that the anticommutator of two such fermionic constraints
%reproduces precisely the relevant parts of the bosonic constraints, and in particular
%the canonical Hamiltonian, including the non-positive terms by which the Hamiltonian
%differs from the constraint obtained by truncating the (quadratic) $\E$ Casimir
%from level $\ell =3$ onwards.
% But let us first discuss possible extensions 
%of (\ref{S}) {\em beyond supergravity}, because these may teach us about how to
%make the $\E$ model compatible with supergravity beyond $\ell =3$.  Sticking with the 
%terms linear in the fermions, one would like to extend the sum over all roots. For the
%timeline imaginary roots of higher multiplicity there is the problem that we do not know
%how to contract fermions with roots of higher multiplicity than the null roots, so let us 
%restrict attention in first approximation to real roots. Then we would have

Since the Weyl group $W(\E)$ of $\E$ can be embedded in $K(\E)$ it would seem like a minimal requirement to extend the expression~\eqref{S} by complete Weyl orbits of roots. As the real roots of $\E$ form a single Weyl orbit ($\E$ is simply-laced) this would lead to the following expression for $\cS_A$:
\begin{align}
\label{S1}
\cS_A &= \pi\!\cdot \!\phi_A\, + \frac12
\sum_{\alpha^2 =2} \Pof{\alpha} \big(\Gt(\alpha) \phi(\alpha)\big)_A  
+ \cdots ,
\nn\\[2mm]
  &\equiv  \pi\!\cdot \!\phi_A \; + \frac12 
\sum_{w\in W(\E)} \Pof{w(\alpha_0)} \big(\Gt(w(\alpha_0)) 
\phi(w(\alpha_0))\big)_A   + \cdots
\end{align}
where the dots now indicate terms associated with imaginary roots. In the second line $\alpha_0$ represents an arbitrary real root. %(as all real roots belong to a single Weyl orbit for $\E$).
%where the sum is now over the Weyl group of $\E$ and $\alpha_0$ is any
%simple root (as for simply laced hyperbolic algebras, the real roots constitute 
%the Weyl orbit of a simple root). 
We see again that the minus sign in (\ref{Gamma1})
is essential, otherwise the contributions from positive and negative roots
would cancel in the sum. As we showed in section~\ref{sec:susytrm}, this expression containing only the real roots is incompatible with $K(\E)$. The expression is, however, compatible with the $\E$ Weyl group. But the anticommutator would now give rise to an infinity of new terms that do not seem to make sense.

We know from supergravity that we require also contributions from null imaginary roots ($\alpha^2=0$) and these would need to be covariantised under the Weyl group as well. We will not investigate the effect of this covariantisation here since already the real roots are problematic. A uniform treatment of all $\E$ roots requires also the inclusion of time-like imaginary roots ($\alpha^2<0$). These come with higher multiplicity and their addition to $\cS_A$ might necessitate higher spin realizations of the type constructed in \cite{Kleinschmidt:2013eka}, so as to be able to contract the relevant polarisation tensors with the fermions.

\subsection{Adding fermions}
\label{sec:ferm}

We now consider some aspects of the inclusion of fermionic degrees of freedom (at lowest order). Let $\Psi$ be a spinorial representation $\Psi$ of the compact subgroup and consider the Lagrangian
\begin{align}
\label{Lfermion}
L = L_B + L_F = \frac12 \langle P| P\rangle - \frac{i}2 \langle \Psi| D\Psi\rangle,
\end{align}
where $D\Psi$ is the $K$-covariant derivative with the composite connection $Q$ constructed out of $V$. In triangular gauge one has $\Qof{\alpha}=\Pof{\alpha}$ for all positive root components.

We can write out the covariant derivative in the vector-spinor representation for $\alpha>0$ (real or imaginary) as follows
\begin{align}
L_F=-\frac{i}{2}\langle \Psi | D\Psi\rangle = -\frac{i}{2}G_{\ta\tb} \phi^\ta \partial \phi^\tb +\frac{i}2 \sum_{\alpha>0}\sum_{r=1}^{\mathrm{mult}(\alpha)} \Pof{\alpha}^r\, \jof{\alpha}^r,
\end{align}
where $\jof{\alpha}^r$ denotes the fermion bilinear constructed out of the action of the $\kof{\alpha}^r$ in the vector-spinor representation and then contracted in the invariant bilinear form: $\jof{\alpha}^r = G_{\ta\tb} \phi^\ta \delta_\alpha^r \phi^\tb=G_{\ta\tb}\phi^\ta(\kof{\alpha}^r\cdot\phi^\tb)=-2\Jof{\alpha}^r$. We will suppress the multiplicity index $r$ in our schematic discussion below in order to avoid cluttering the expressions.

The canonical fermionic momentum from~\eqref{Lfermion} is 
\begin{align}
\varpi_\ta= \frac{\partial^L L}{\partial \partial\phi^\ta} = \frac{i}2G_{\ta\tb} \phi^\tb,
\end{align}
where we are using left Grassmann derivatives. The momentum satisfies the Poisson bracket
\begin{align}
\ld \phi^\ta , \varpi_\tb \rd =-1.
\end{align}
The corresponding (classical) Dirac bracket is therefore %(MISSING FACTOR 1/2???)
\begin{align}
\ld \phi^\ta, \phi^\tb \rd = i G^{\ta\tb}.
\end{align}
Above we were using this bracket without the factor of $i$ by thinking of the $\phi^\ta$ as quantum operators. The additional $i$ here implies that at the classical level 
\begin{align}
\label{jj}
\ld \jof{\alpha},\jof{\beta} \rd =-2 i (c_{\alpha,\beta}\, \jof{\alpha+\beta}-c_{\alpha,-\beta}\, \jof{\alpha-\beta}).
\end{align}

Let us denote the bosonic conjugate momenta in the theory with fermions by $\hat\Pi$. Then we get
\begin{align}
\pihat_\ta &= \frac{\partial L}{\partial \partial q^\ta}=G_{\ta\tb}\partial q^\tb = \pi_\ta,\\
\Pihof{\alpha} &= \frac{\partial L}{\partial \partial \Aof{\alpha}}=\sum_{\beta>0} \left(2\Pof{\beta}+\frac{i}2 \jof{\beta}\right)\frac{\partial \Pof{\beta}}{\partial\partial \Aof{\alpha}}.
\end{align}
We note that the momenta conjugate to the Cartan subalgebra variables $q^\ta$ do not change (since these do not couple to the fermions) and that the matrix $\frac{\partial \Pof{\beta}}{\partial\partial \Aof{\alpha}}$ relating the conjugate momenta to the  $\Pof{\beta}$ is identical to the purely bosonic theory. This means that the inversion proceeds in exactly the same way as in~\eqref{pi2p}, leading to
\begin{align}
\Pof{\alpha}+\frac{i}4 \jof{\alpha} = e^{-q^\ta\alpha_\ta} \left(\Pihof{\alpha}-\frac12\sum_\beta c_{\beta,\alpha} \Aof{\beta}\Pihof{\alpha+\beta} + \ldots\right).
\end{align}
We now introduce the notation 
\begin{align}
\Phof{\alpha} \equiv \Pof{\alpha} +\frac{i}4 \jof{\alpha}.
\end{align}
Since $\Pihof{\alpha}$ and $\Aof{\alpha}$ are conjugate variables as before, we deduce that we have the following canonical commutation relations
\begin{align}
\label{PhPh}
\ld \Phof{\alpha} , \Phof{\beta}\rd &= c_{\alpha,\beta} \Phof{\alpha+\beta},\nn\\
\ld \pihat_\ta,\Phof{\alpha} \rd &= \alpha_\ta \Phof{\alpha}.
\end{align}
The new `supercovariant' velocity components $\Phof{\alpha}$ therefore satisfy the same Borel algebra as the $\Pof{\alpha}$ in the purely bosonic theory. In terms of the original velocities, and in view of~\eqref{jj}, one therefore has
\begin{align}
\ld \Pof{\alpha}, \Pof{\beta} \rd &= c_{\alpha,\beta}\Phof{\alpha+\beta} - \frac{i}{8} c_{\alpha,\beta} \jof{\alpha+\beta}+\frac{i}{8} c_{\alpha,-\beta} \jof{\alpha-\beta},\nn\\
\ld \pi_\ta, \Pof{\alpha}\rd &%= \alpha_\ta \left(\Pof{\alpha}+\frac{i}{2} \jof{\alpha}\right)
= \alpha_\ta \Phof{\alpha}.
\end{align}
The appearance of $\Phof{\alpha}$ on the r.h.s. in these equations is important.
For deriving this, we used that $\Pof{\alpha}$ and $\jof{\alpha}$ commute while $\Phof{\alpha}$ and $\jof{\alpha}$ satisfy
\begin{align}
\label{Phj}
\ld \Phof{\alpha} , \jof{\beta} \rd =  \ld \frac{i}4  \jof{\alpha},\jof{\beta} \rd = \frac{1}{2} c_{\alpha,\beta} \jof{\alpha+\beta} -\frac{1}{2} c_{\alpha,-\beta} \jof{\alpha-\beta}.
\end{align}

Let us verify the consistency of the relation~\eqref{PhPh} and~\eqref{Phj} in the equations of motion. In the model~\eqref{Lfermion} one has on the one hand the Euler--Lagrange equations
\begin{align}
\partial \Pof{\alpha} &= 
-\pi^\ta \alpha_\ta \Pof{\alpha}+2\sum_{\beta>0} 
c_{\alpha,\beta} \Pof{\beta}\Pof{\alpha+\beta} -\frac{i}4 (\alpha\cdot \pi) \jof{\alpha}
+\frac{i}4 \sum_{\beta>0}\Pof{\beta} \left( c_{\alpha,\beta}\jof{\alpha+\beta} + c_{\alpha,-\beta} \jof{\alpha-\beta}\right)\nn\\
&= -(\alpha\cdot \pi) \Phof{\alpha} + 2\sum_{\beta>0} c_{\alpha,\beta}\Pof{\beta} \Phof{\alpha+\beta} -\frac{i}{4} \sum_{\beta>0} \Pof{\beta}\left(c_{\alpha,\beta}  \jof{\alpha+\beta}-c_{\alpha,-\beta}  \jof{\alpha-\beta}\right)
\end{align}
The Hamiltonian on the other hand is (as before)
\begin{align}
H = \frac12\langle P | P\rangle = \frac12 \pihat_\ta G^{\ta\tb} \pihat_\tb + \sum_{\alpha>0} \Pof{\alpha} \Pof{\alpha},
\end{align}
in terms of the `old' purely bosonic $P$.\footnote{That this is true can be seen in the following simple example involving a derivative coupling. Let $L=\frac12 \dot{q}^2 + \dot{q} j$. The conjugate momentum is $\hat{p}=\dot{q}+j\equiv p+j$ and the Hamiltonian is $H=\hat{p}\dot{q} -L = \frac12\dot{q}^2=\frac12p^2$.}
The Hamiltonian equations of motion for $\Pof{\alpha}$ are then
\begin{align}
\partial \Pof{\alpha} &= \ld \Pof{\alpha}, H \rd = -\pi_\ta \ld \pi^\ta, \Pof{\alpha} \rd + 2 \sum_{\beta>0}\Pof{\beta} \ld \Pof{\alpha},\Pof{\beta} \rd\nn\\
&= -(\alpha\cdot\pi) \Phof{\alpha} +2\sum_{\beta>0} c_{\alpha,\beta} \Pof{\beta}\Phof{\alpha+\beta} -\frac{i}{4} \sum_{\beta>0} \Pof{\beta}\left(c_{\alpha,\beta}  \jof{\alpha+\beta}-c_{\alpha,-\beta}  \jof{\alpha-\beta}\right)
\end{align}
in complete agreement with the Lagrangian equations.

\subsection{Final comments}

The underlying problems of the non $K(\E)$ covariance of the supersymmetry constraint $\cS$ appears to be the unfaithfulness of the spinor representation that was used to construct $\cS$. A full understanding of this issue  requires a more detailed understanding of the representation theory of $K(\E)$. This does not only involve the construction of faithful fermionic representations but also a study of the properties of the `coset representation' $P$ and the decomposition of its tensor products with fermionic representations. 

Finding a supersymmetric $\E$ model might exhibit a feature similar to one of the hallmarks of superstring theory. In superstring theory, supersymmetry is implemented only on the two-dimensional worldsheet but the consistency conditions of the theory imply that there is also supersymmetry in the target space-time, leading to supergravity at low energies. It is not inconceivable that a supersymmetric $\E$ model on a worldline would similarly induce supersymmetry in the algebraically generated space-time. The close connection between the fermionic $\E$ model on the worldline and the space-time supergravity equations found in~\cite{Damour:2005zs,deBuyl:2005mt,Damour:2006xu} could be viewed as evidence for this idea.

The problem of finding a $K(\E)$ covariant supersymmetry constraint~\eqref{PodotPsi} can be phrased representation theoretically as follows. Both the coset velocity $P$ and the vector-spinor $\Psi$ are honest $K(\E)$ representations. Their tensor product $P \otimes \Psi$ is also a $K(\E)$ representation and the question is what the invariant subspaces of this tensor product are; in particular, if there is a Dirac-spinor representation $\mathcal{S}$ contained in it. To the best of our knowledge very little is known about these kinds of questions since $K(\E)$ is not a Kac--Moody algebra. Already the decomposability (or not) of the coset velocity $P$ itself is an open question. 
If $P$ {\em was} decomposable this could have important consequences for the 
construction of invariant Lagrangians.

Finally, we note that similar issues already arise for the affine case \cite{NS1,NS2}, where
$K(\E)$ is replaced by the simpler (but still infinite-dimensional) involutory
subgroup $K({\rm E}_9)\subset {\rm E}_9$. In 
that case one is dealing with a {\em field theory} in two dimensions, rather than a worldline 
model, and the faithfulness of the  $K({\rm E}_9)$ representations is ensured {\em on-shell} 
by the additional dependence on the space coordinate and the differential relations
obeyed by the transformation coefficients. For the {\em off-shell} theory, however,  the
existence and construction of faithful representations remains an open problem there
as well.

\vspace{7mm}
\noindent
\textbf{Acknowledgements:} 
We thank Thibault Damour, Fran\c{c}ois Englert and Marc Henneaux for discussions related to
this work, and Ralf K\"ohl for correspondence on parametrisations of the unipotent 
group $N$ in the Kac--Moody case.

%%%%%%%%%%%%%%%%%%%%%%%%

%\newpage

\appendix

\section{The second quantised vector-spinor for imaginary roots}
\label{app:vsrep}

The Dirac-spinor representation is insensitive to the polarisation (i.e., multiplicity) of the imaginary roots but the more faithful vector-spinor representation can be used to derive partial information on the structure constants of $K(\E)$. The more faithful higher-spin realisations of~\cite{Kleinschmidt:2013eka} in principle capture even more information on the imaginary roots.

In this appendix, we study in more detail the representation of the vector-spinor that was completely determined by its values on the real roots by~\eqref{Jreal}. For any real root $\alpha$ of $\E$ we recall that the canonical $K(\E)$ generators are given by
\begin{align}
\label{Jreal2}
\kof{\alpha} = X_{\ta\tb}(\alpha) \phi^\ta \Gt(\alpha) \phi^\tb,\quad
\textrm{with $X_{\ta\tb}(\alpha) = -\frac12 \alpha_\ta\alpha_\tb +\frac14 G_{\ta\tb}$.}
\end{align}
We seek to obtain similar general expression for null roots $\delta$, satisfying $\delta^2=0$ and timelike root $\Lambda$ with $\Lambda^2=-2$.

\subsection*{Null roots $\delta$}

All null roots $\delta$ of $\E$ have multiplicity $\mathrm{mult}(\delta)=8$ and we therefore require the representation of eight generators $\kof{\alpha}^r$. To arrive at the expression, we decompose $\delta=\alpha+(\delta-\alpha)$ for a {\em real} root $\alpha$. Then $(\delta-\alpha)$ is also real and $\delta\cdot\alpha=0$. Employing then the commutator
\begin{align}
\lb \kof{\alpha}, \kof{\delta-\alpha} \rb =\vepst_{\alpha,\delta-\alpha} \kof{\delta}^{(\alpha)},
\end{align}
where we have indicated that there are different possibilities for $\kof{\delta}^{(\alpha)}$. One knows a priori that there are at most eight independent generators.

Substituting in the explicit expression for the real root generators~\eqref{Jreal2} one finds that in the (second quantised) vector-spinor representation
\begin{align}
\kof{\delta}^{(\alpha)} = -2\alpha_{[\ta} \delta_{\tb]} \phi^\ta \Gt(\delta) \phi^\tb,
\end{align}
where $\alpha\cdot \delta=0$. To bring this into a form that brings out the multiplicity $\mathrm{mult}(\delta)=8$, we note that shifting $\alpha\to\alpha+\delta$ does not change the expression, so that we can also summarise it by
\begin{align}
\label{Jnull}
\kof{\delta}^r = \epsilon^r_{[\ta} \delta_{\tb]} \phi^\ta \Gamma(\delta) \phi^\tb,
\end{align}
where the `polarisation vector' $\epsilon^r$ is orthogonal (transverse) to $\delta$ in the DeWitt metric and there is also a gauge invariance $\epsilon^r\to \epsilon^r+\delta$. This leaves eight independent choices which agrees with the multiplicity of the null root of $\E$. Note that the transversality of the polarisation vector is \textit{not} manifest in~\eqref{Jnull};  it is rather a consequence of the way the generator is constructed from commutators of real roots.

%One choice of labelling of the null roots can be given by nine-tuples of indices, motivated by $E^{a_0|a_1\ldots a_8}$ on level 3 in the $SL(10)$-decomposition of $\E$. The null roots are those where all nine indices are different but one has to take into account the irreducibility constraint
%\begin{align}
%\label{irred}
%E^{[a_0|a_1\ldots a_8]}=0.
%\end{align}
%This could be used for example to never have the smallest index in front. Then for the index choice $(a_0, a_1,\ldots,a_8)=(1,2,\ldots,9)$ there are eight choices, agreeing with the multiplicity.

\subsection*{Imaginary roots $\Lambda^2=-2$}

It is also possible to derive the general form of the $\Lambda^2=-2$ generators from commuting two real root generators in a way similar to above. Let $\Lambda=\alpha+(\Lambda-\alpha)$ with $\alpha^2=(\Lambda-\alpha)^2=2$, then  $\alpha\cdot\Lambda =-1$. We know that
\begin{align}
\left[ \kof{\alpha}, \kof{\beta} \right] = \vepst_{\alpha,\beta} \kof{\Lambda}^{(\alpha)}.
\end{align}
By substituting in the explicit form for the real root generators one finds
\begin{align}
\kof{\Lambda}^{(\alpha)} = 2 Y_{\ta\tb}(\alpha) \phi^\ta \Gamma(\Lambda) \phi^\tb
\end{align}
with
\begin{align}
\label{Ytens}
Y_{\ta\tb}(\alpha)= Y_{\tb\ta}(\alpha)&=
-\alpha_{(\ta} \Lambda_{\tb)} + \alpha_\ta \alpha_\tb -
\frac14 \Lambda_\ta \Lambda_\tb +\frac18 G_{\ta\tb} \nn\\
&= v_{(\ta} \Lambda_{\tb)} +a_{\ta\tb}
\end{align}
for
\begin{align}
v_\ta &= -\alpha_\ta -\frac58 \Lambda_\ta,\\
a_{\ta\tb} &= \alpha_\ta\alpha_\tb +\frac38 \Lambda_\ta\Lambda_\tb +\frac18 G_{\ta\tb}.
\end{align}
The separation of the $\Lambda_{(a} \Lambda_{b)}$ here was chosen in such a way that
\begin{align}
\Lambda^\tb a_{\tb\ta} = v_\ta
\end{align}
and is motivated by vertex operator algebra (VOA) constructions. The gauge symmetries of the parametrisation~\eqref{Ytens} are
\begin{align}
a_{\ta\tb} &\to a_{\ta\tb} + 2\epsilon_{(\ta} \Lambda_{\tb)}\\
v_\ta &\to v_\ta - 2\epsilon_\ta
\end{align}
and also leave the above condition $\Lambda^\tb a_{\tb\ta} = v_\ta$ invariant. 
The parameter $\epsilon$ here is chosen orthogonal to $\Lambda$. We also note that 
$\Lambda^\ta\Lambda^\tb Y_{\ta\tb}=19/4$ and $G^{\ta\tb} Y_{\ta\tb}=-9/4$ are gauge 
invariant and constrain the tensor $Y_{\ta\tb}$. The count is

\vspace{5mm}

\begin{tabular}{c|l}
55 & components of $a_{\ta\tb}$\\
-9 & components of $\epsilon$ such that $\epsilon\cdot\Lambda=0$\\
-2 & norm conditions on $Y_{\ta\tb}$\\
\hline
44 & multiplicity of root space
\end{tabular}

\vspace{5mm}

This count does not completely parallel the VOA construction and it would be desirable to have an interpretation in terms of Young symmetries similar to the null case above.

\subsection*{Conjectural form for any generator}

Similar to the formula for roots satisfying $\Lambda^2=-2$ as above, we can give a tentative form of the action of any generator $\kof{\Lambda}^r$ on the vector-spinor $\phi^\ta$ for an arbitrary imaginary root $\Lambda$. This form rests on the assumption that any generator in the root space of $\Lambda$ can be written as the commutator of two {\em real} root generators, that is
\begin{align}
 c_{\alpha,\beta} \kof{\Lambda}^{(\alpha)} = \lb \kof{\alpha},\kof{\beta}\rb
\end{align}
for $\alpha,\beta>0$ and real with $\alpha+\beta=\Lambda$. (We note that there the second term in the commutation relation~\eqref{kcommutator} vanishes automatically for the configuration chosen here since $\alpha-\beta$ is not a root.) The conjecture is that as $\alpha$ and $\beta$ traverse all possible decompositions of $\Lambda$, their commutators contain a basis of the root space of $\Lambda$. The number of  decompositions of $\Lambda$ is larger than $\mathrm{mult}(\Lambda)$ and many of the commutators will be linearly dependent. What we require is that the space generated  by all possible commutators is equal to the full root space of $\Lambda$:
\begin{align}
\left.\bigg\langle \kof{\Lambda}^{(\alpha)}\,\middle|\, \alpha>0,\, \alpha^2=2,\, (\Lambda-\alpha)^2=2 \right.\bigg\rangle
=\left.\bigg\langle \kof{\Lambda}^r\,\middle|\, r=1,\ldots,\mathrm{mult}(\Lambda) \bigg\rangle\right.
\end{align}
This is a stronger version of a conjecture already contained~\cite{Brown:2004jb} which only addressed the decomposition of the imaginary root vector $\Lambda=\alpha+\beta$ into two real roots. In all cases that we checked the assumption we are making is true but we are not aware of a general proof.

Under this assumption, we can find the following formula for $\kof{\Lambda}^{(\alpha)}$ {\em in the vector-spinor representation}. We have to distinguish the cases $\Lambda^2=-4k$ and $\Lambda^2=2-4k$ for integer $k\geq 0$ because of the (anti-)symmetry of $\Gt(\Lambda)$. The condition that $\alpha$ and $\Lambda-\alpha$ be real implies that $\alpha\cdot(\Lambda-\alpha)=\frac12(\Lambda^2-4)$. Then the calculations are completely analogous to the two cases described above, leading to% (CHECK)
\begin{subequations}
\begin{align}
\Lambda^2&=-4k &:&& \kof{\Lambda}^{(\alpha)} &= 2(k-1)\alpha_{[\ta} \Lambda_{\tb]} \phi^\ta \Gt(\Lambda)\phi^\tb ,\\
\Lambda^2&=2-4k &:&& \kof{\Lambda}^{(\alpha)} &= \left(-2k \alpha_{(\ta} \Lambda_{\tb)}+2k\alpha_\ta\alpha_\tb -\frac12\Lambda_\ta\Lambda_\tb +\frac14 G_{\ta\tb} \right)\phi^\ta \Gt(\Lambda)\phi^\tb .
\end{align}
\end{subequations}
These expressions were derived under the assumption that $c_{\alpha,\Lambda-\alpha}=-\vepst_{\alpha,\Lambda-\alpha}$.

\section{Some more explicit results for $GL(3,\reals)/SO(3)$}
\label{app:GL3}

For $GL(3,\reals)$ the coset element is
\begin{align}
V=AN= \begin{pmatrix}
e^{q^1}&&\\
&e^{q^2}&\\
&&e^{q^3} 
\end{pmatrix}
\begin{pmatrix}
1 & N^1{}_\ttwo & N^1{}_\tthree\\
&1 & N^2{}_\tthree\\
&&1
\end{pmatrix}.
\end{align}
The notation here is such that an index value with a tilde refers to a curved (world) index and an index value without a tilde to a flat (tangent space) direction.
The inverse of $N$ is given by
\begin{align}
N^{-1} = \begin{pmatrix}
1 & N^\tone{}_2 & N^\tone{}_3\\
&1 & N^\ttwo{}_3\\&&1
\end{pmatrix}
=\begin{pmatrix}
1 & -N^1{}_\ttwo & -N^1{}_\tthree + N^1{}_\ttwo N^2{}_\tthree\\
&1&-N^2{}_\tthree\\
&&1
\end{pmatrix}.
\end{align}
With this parametrisation it is straightforward to compute the 
coset velocity from $\partial V V^{-1}$,
\begin{align}
P &= \begin{pmatrix}
P_{11} & P_{12} & P_{13}\\
P_{21} & P_{22} & P_{23}\\
P_{31} & P_{32} & P_{33}
\end{pmatrix}\nn\\
&=\begin{pmatrix}
\partial{q}^1 & \frac12 e^{q^1-q^2} \partial{N}^1{}_\ttwo & \frac12 e^{q^1-q^3} \left(\partial{N}^1{}_\tthree +\partial{N}^1{}_\ttwo N^\ttwo{}_3\right)\\
\frac12e^{q^1-q^2}\partial{N}^1{}_\ttwo & \partial{q}^2 & \frac12 e^{q^2-q^3} \partial{N}^2{}_\tthree\\
\frac12 e^{q^1-q^3} \left(\partial{N}^1{}_\tthree +\partial{N}^1{}_\ttwo N^\ttwo{}_3\right)& \frac12 e^{q^2-q^3} \partial{N}^2{}_\tthree &  \partial{q}^3
\end{pmatrix}
\end{align}
where, of course, $P_{ab} = P_{ba}$.

The bosonic Lagrangian is
\begin{align}
L &= \frac12 \big[\Tr (P^2) - (\Tr P)^2\big]\nn\\[2mm]
&=  \frac12 G_{\ta\tb} \partial{q}^\ta \partial{q}^\tb + P_{12}^2 + P_{13}^2 + P_{23}^2\nn\\[2mm]
& = \frac12 G_{\ta\tb} \partial{q}^\ta \partial{q}^\tb
+ \frac14 e^{2q^1-2q^2} \left(\partial{N}^1{}_\ttwo \right)^2+ \frac14 e^{2q^2-2q^3} \left(\partial{N}^2{}_\tthree \right)^2+ \frac14 e^{2q^1-2q^3} \left(\partial{N}^1{}_\tthree+\partial{N}^1{}_\ttwo N^\ttwo{}_3\right)^2,
\end{align}
where $\frac12 G_{\ta\tb} \partial{q}^\ta \partial{q}^\tb= -\partial{q}^1\partial{q}^2 -\partial{q}^1\partial{q}^3 -\partial{q}^2\partial{q}^3$.
The conjugate momenta are
\begin{align}
\label{conm}
\pi_\ta &= G_{\ta\tb} \partial{q}^\tb,\nn\\[2mm]
\Pi^\ttwo{}_1 &= \frac{\partial L}{\partial \partial{N}^1{}_\ttwo} = \frac12 e^{2q^1-2q^2} \partial{N}^1{}_\ttwo-\frac12e^{2q^1-2q^3} N^2{}_\tthree \left(\partial{N}^1{}_\tthree+\partial{N}^1{}_\ttwo N^\ttwo{}_3\right)\nn\\[2mm]
&= e^{q^1-q^2} P_{21} + e^{q^1-q^3} N^\ttwo{}_3 P_{31},\\[2mm]
\Pi^\tthree{}_1 &= \frac{\partial L}{\partial \partial{N}^1{}_\tthree} = \frac12 e^{2q^1-2q^3}  \left(\partial{N}^1{}_\tthree+\partial{N}^1{}_\ttwo N^\ttwo{}_3\right) 
= e^{q^1-q^3} P_{31}\\
\Pi^\tthree{}_2 &= \frac{\partial L}{\partial \partial{N}^2{}_\tthree} = \frac12e^{2q^2-2q^3} \partial{N}^2{}_\tthree
= e^{q^2-q^3} P_{32}.
\end{align}
These relations can be inverted to give $\partial{q}^\ta = G^{\ta\tb} \pi_\tb$ and
\begin{align}
\partial{N}^1{}_\ttwo &= 2 e^{-2q^1+2q^2} \big(\Pi^\ttwo{}_1+ N^2{}_\tthree \Pi^\tthree{}_1\big)
\nn\\[2mm]
\partial{N}^1{}_\tthree &= 2e^{-2q^1+2q^3} \Pi^\tthree{}_1 + 2e^{-2q^1+2q^2} N^2{}_\tthree \big(\Pi^\ttwo{}_1+ N^2{}_\tthree \Pi^\tthree{}_1\big)\nn\\[2mm]
\partial{N}^2{}_\tthree &= 2 e^{-2q^2+2q^3} \Pi^\tthree{}_2.
\end{align}
Equivalently,
\begin{align}
\label{invrels}
P_{12} &= e^{-q^1+q^2} \big(\Pi^\ttwo{}_1 + N^2{}_\tthree \Pi^\tthree{}_1\big),\nn\\
P_{23} &= e^{-q^2+q^3} \Pi^\tthree{}_2,\nn\\
P_{13} &= e^{-q^1+q^3} \Pi^\tthree{}_1,
\end{align}
in agreement with the general formula~\eqref{Pab}, up to an overall factor. Let us denote
\begin{align}
P(\alpha_{12}) \equiv P_{12}\, ,\quad
P(\alpha_{23}) \equiv P_{23}\, , \quad
P(\alpha_{13}) \equiv P_{13},\quad
\end{align}
and
\begin{align}
\alpha_{12} = (1,-1,0),\quad
\alpha_{13} = (1,0,-1),\quad
\alpha_{23} = (0,1,-1).
\end{align}
Then the  Hamiltonian is
\begin{align}
H&= \frac12 \pi_\ta G^{\ta\tb} \pi_\tb+ e^{-2q^1+2q^2} 
\big(\Pi^\ttwo{}_1 + N^2{}_\tthree \Pi^\tthree{}_1\big)^2 + e^{-2q^1+2q^3}  
\big(\Pi^\tthree{}_1\big)^2 +  e^{-2q^2+2q^3} \big(\Pi^\tthree{}_2\big)^2\nn\\[2mm]
&=\frac12 \pi_\ta G^{\ta\tb} \pi_\tb + \sum_{a<b} P(\alpha_{ab})^2.
\end{align}

Using the canonical brackets $\left\{q,p\right\}=1$ between the conjugate 
variables we recover the relations already previously derived (for $a<b$ and any $\alpha>0$)
\begin{align}
\left\{ \pi_\ta,\pi_\tb\right\} &= 0,\nn\\[2mm]
\left\{ \pi_\ta,P(\alpha)\right\} &= \alpha_\ta P(\alpha),\nn\\[2mm]
\left\{ P(\alpha_{ab}),P(\alpha_{cd})\right\} &= \left\{\begin{array}{cl}
\epsilon_{\alpha_{ab},\alpha_{cd}} P(\alpha_{ab}+\alpha_{cd}) & \textrm{if $\alpha_{ab}+\alpha_{cd}$ is a root,}\\
0 & \textrm{otherwise.}
\end{array}\right.
\end{align}

The conserved current is
\begin{align}
J = V^{-1} P V = \begin{pmatrix}
J^\tone{}_\tone & J^\tone{}_\ttwo & J^\tone{}_\tthree\\
J^\ttwo{}_\tone & J^\ttwo{}_\ttwo & J^\ttwo{}_\tthree\\
J^\tthree{}_\tone & J^\tthree{}_\ttwo & J^\tthree{}_\tthree
\end{pmatrix}
\end{align}
with
\begin{align}
J^\ttwo{}_\tone &= \Pi^\ttwo{}_1,\nn\\
J^\tthree{}_\tone &= \Pi^\tthree{}_1,\nn\\
J^\tthree{}_\ttwo &= \Pi^\tthree{}_2+ N^1{}_\ttwo\Pi^\tthree{}_1 ,
\end{align}
below the diagonal and the following diagonal and upper triangular components
\begin{align}
J^\tone{}_\tone &= G^{1\ta}\pi_\ta - N^1{}_\ttwo \Pi^\ttwo{}_1 - N^1{}_\tthree \Pi^\tthree{}_1\nn\\[1mm]
J^\ttwo{}_\ttwo &= G^{2\ta}\pi_\ta + N^1{}_\ttwo \Pi^\ttwo{}_1 - N^2{}_\tthree \Pi^\tthree{}_2\nn\\[1mm]
J^\tthree{}_\tthree &= G^{3\ta}\pi_\ta + N^1{}_\tthree \Pi^\tthree{}_1 + N^2{}_\tthree \Pi^\tthree{}_2\\[1mm]
J^\tone{}_\ttwo &= e^{-2q^1+2q^2} (\Pi^\ttwo{}_1 + N^2{}_\tthree \Pi^\tthree{}_1) + 
N^1{}_\ttwo\left(\pi_\tone -\pi_\ttwo - N^1{}_\tthree\Pi^\tthree{}_1+ N^2{}_\tthree\Pi^\tthree{}_2\right) - N^1{}_\tthree \Pi^\tthree{}_2 - N^1{}_\ttwo N^1{}_\ttwo \Pi^\ttwo{}_1\nn\\[1mm]
J^\tone{}_\tthree &= e^{-2q^1+2q^3} \Pi^\tthree{}_1 + e^{-2q^1+2q^2}N^2{}_\tthree(\Pi^\ttwo{}_1 + N^2{}_\tthree \Pi^\tthree{}_1) -e^{-2q^2+2q^3} N^1{}_\ttwo \Pi^\tthree{}_2\nn\\[1mm]
&\quad + N^1{}_\tthree \left( \pi_\tone -\pi_\tthree - N^1{}_\ttwo \Pi^\ttwo{}_2- N^2{}_\tthree \Pi^\tthree{}_2\right)
 - N^1{}_\ttwo N^2{}_\tthree \left(\pi_\ttwo - \pi_\tthree-N^2{}_\tthree \Pi^\tthree{}_2\right) - N^1{}_\tthree N^1{}_\tthree \Pi^\tthree{}_1\nn\\[1mm]
 J^\ttwo{}_\tthree &= e^{-2q^2+2q^3} \Pi^\tthree{}_2 + N^2{}_\tthree(\pi_\ttwo-\pi_\tthree) + N^1{}_\tthree \Pi^\ttwo{}_1 - N^2{}_\tthree N^2{}_\tthree \Pi^\tthree{}_2.\nn
\end{align}
The relation of the components of the conserved charge to the canonical momenta was already discussed in~\cite{Damour:2002et}. The `lowest' components of $J$ are just identical to the canonical momenta and the structure gets increasingly complicated for higher and higher components. For infinite-dimensional algebras (without a lowest component) this description breaks down without a suitable truncation.

One can now check that
\begin{align}
H = \frac12 \big[\Tr (J^2) - (\Tr J)^2\big]\nn\\
\end{align}
and that the components of the current satisfy the $GL(3)$ algebra,
\begin{align}
\left\{ J^i{}_j, J^k{}_l \right\} = \delta^k_j J^i{}_l - \delta^i_l J^k{}_j.
\end{align}

\end{document}